# Low-cost machine learning approach to the prediction of transition metal phosphor excited state properties

Gianmarco Terrones[1], Chenru Duan[1,2], Aditya Nandy[1,2], and Heather J. Kulik[1,2]

[1]*Department of Chemical Engineering, Massachusetts Institute of Technology, Cambridge, MA 02139*

[2]*Department of Chemistry, Massachusetts Institute of Technology, Cambridge, MA 02139*

ABSTRACT: Photoactive iridium complexes are of broad interest due to their applications ranging from lighting to photocatalysis. However, the excited state property prediction of these complexes challenges *ab initio* methods such as time-dependent density functional theory (TDDFT) both from an accuracy and a computational cost perspective, complicating high throughput virtual screening (HTVS). We instead leverage low-cost machine learning (ML) models to predict the excited state properties of photoactive iridium complexes. We use experimental data of 1,380 iridium complexes to train and evaluate the ML models and identify the best-performing and most transferable models to be those trained on electronic structure features from low-cost density functional theory tight binding calculations. Using these models, we predict the three excited state properties considered, mean emission energy of phosphorescence, excited state lifetime, and emission spectral integral, with accuracy competitive with or superseding TDDFT. We conduct feature importance analysis to identify which iridium complex attributes govern excited state properties and we validate these trends with explicit examples. As a demonstration of how our ML models can be used for HTVS and the acceleration of chemical discovery, we curate a set of novel hypothetical iridium complexes and identify promising ligands for the design of new phosphors.

# 1. Introduction.

The interactions between light and matter underpin phenomena ranging from photovoltaics[1] to photosynthesis[2] to bioluminescence.[3] The design of new functional materials that can leverage light-matter interactions has led to significant technological advancements.[4-6] Exemplary of these advancements are iridium photoactive complexes that have been investigated extensively due to their applications in lighting and display technology[7-10], photocatalysis[11-13], and bioimaging.[14,15] The spin-orbit coupling (SOC) produced by iridium causes these complexes to efficiently convert excitons into light or chemical energy.[16] Simultaneously, iridium uniquely limits nonradiative decay rates by destabilizing a metal-centered ($^3$MC) triplet excited state due to strong metal-ligand bonding[17], further improving complex efficiency. In these iridium-centered complexes, the judicious selection of ligands allows for the modulation of phosphorescence color (i.e., emission energy) and efficiency/brightness (i.e., via modulating excited state lifetime and photoluminescence quantum yield). The accurate prediction of these three excited state properties is important for the discovery of novel iridium complexes to enable vibrant display technologies and green photocatalysis.

To screen a large number of compounds, computational modeling with time-dependent density functional theory (TDDFT) can enable rapid transition metal complex property prediction. While these methods are commonly employed to estimate emission energies[18-26], the calculation of lifetime and quantum yield is more challenging both from an accuracy and a computational cost perspective. The calculation of lifetime[21,27-31] requires the inclusion of SOC in TDDFT to estimate the transition dipole moment between the $T_1$ sublevels and $S_0$. The calculation of photoluminescence quantum yield further requires the calculation of nonradiative rates, which entails the use of the thermal vibration correlation function rate theory[32-34] and excited state geometry optimization.[34,35] Thus, while *ab initio* computational methods have provided valuable insight into the properties of iridium complexes, they are computation-intensive, requiring around one day of computation time per complex for the least demanding calculations, and may not reach the accuracy required to enable rational design.

Supervised machine learning (ML) has emerged as a powerful complement to *ab initio* methods in recent years due to its capacity to reproduce *ab initio* results at significantly lower cost[36-40], enabling screening vast regions of chemical space.[41] Furthermore, ML models can be trained on experimental data, enabling the prediction of properties that challenge *ab initio* methods, such as material stability.[42] With regard to excited state properties, ML models have been successfully applied for the prediction of phosphorescence energies[43], fluorescence rates[44], and fluorescence energies and quantum yields[45] after training on *ab initio* or experimental data. However, to our knowledge ML has yet to be applied to predict the excited state properties of transition metal (e.g., iridium) photoactive complexes, which are workhorses of photocatalysis and OLED technology. Thus, the need to predict for and simultaneously optimize multiple properties challenging to predict with TDDFT in iridium complexes motivates the application of ML.

In this work, we use ML to predict three key properties of iridium complexes: $Em_{50/50}$ (mean emission energy), excited state lifetime, and emission spectral integral (brightness). We train and evaluate artificial neural networks (ANNs) on a recent experimental dataset[46] of 1,380 iridium(III) phosphors and their properties. We show that features generated with density functional tight binding lead to the most predictive ANN performance and generalization on out-of-sample data. Using these features, we identify trends in phosphor properties, and we extend our models to a new set of hypothetical iridium phosphors. These experimentally-informed ANNs enable fast, accurate prediction of iridium phosphor properties for the rapid exploration of chemical space.

## 2. Data and Representations.

### 2a. Dataset

We constructed the structures of bidentate ligands used in the experimental study of DiLuzio *et al.*[46] on Ir(III) complexes of the form [Ir(**CN**)$_2$(**NN**)]$^+$ (Figure 1). We assigned each ligand as either cyclometalating (**CN**) or ancillary (**NN**), as determined by the two iridium-coordinating atom identities. We studied the same 60 **CN** ligands and 23 **NN** ligands from the prior experimental study[46], excluding only the monodentate DMSO ligand in the prior work, giving rise to a combinatorial set of 1,380 [Ir(**CN**)$_2$(**NN**)]$^+$ phosphor complexes. This set of ligands will be referred to as the high-throughput ligand set (HLS), and we use the same labeling as in the prior study when referring to individual ligands (ESI Tables S1 and S2). We used experimental data from the prior study[46] on the three target properties, Em$_{50/50}$, excited state lifetime, and emission spectral integral. The experimental values for these properties were reported for each of the 1,380 iridium phosphors, and were used for ML model training and performance assessment (ESI Figure S1). **CN** ligands were generated in their neutral form (i.e., with a proton added) when featurized. Because of this, all ligands are neutral with the exception of three **NN** ligands (ESI Text S1). After ligand construction using the draw tool in Avogadro v1.1.2[47,48] and force field (i.e., UFF) optimization, we used these ligands to generate the structures of all possible iridium complexes with one distinct type of **CN** ligand and **NN** ligand using molSimplify v1.6.0[49,50] and force field optimized again.

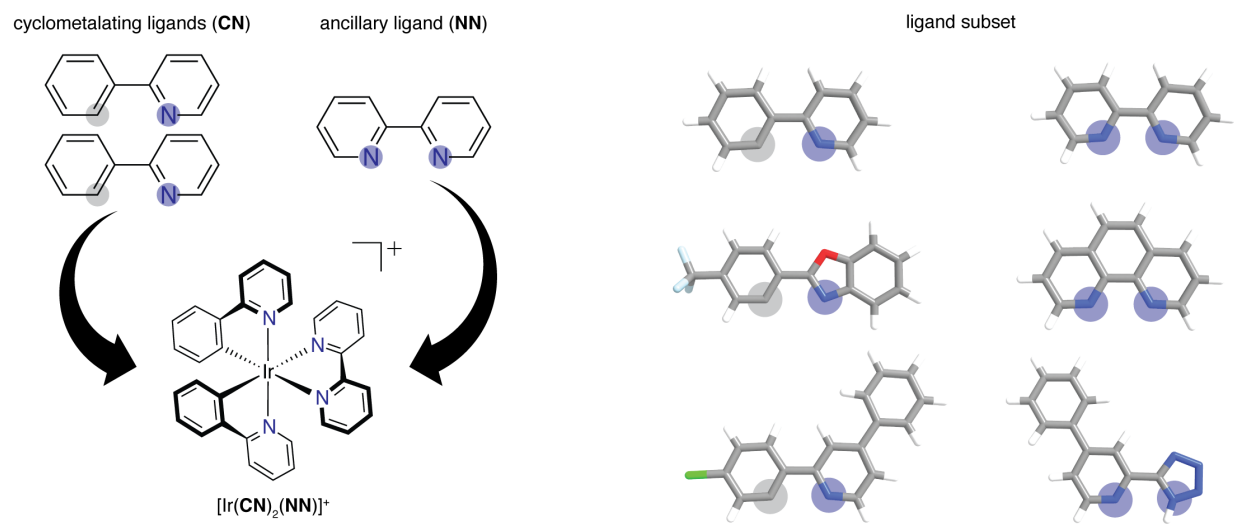

**Figure 1.** (Left) Schematic of how two identical **CN** ligands and one **NN** ligand comprise each of the iridium phosphors studied in this work. Coordinated nitrogen (carbon) atoms are indicated with blue (gray) circles. (Right) Examples of **CN** and **NN** ligands in the experimental dataset of 1,380 iridium phosphors. Atoms are colored as follows: white for hydrogen, gray for carbon, blue for nitrogen, red for oxygen, light blue for fluorine, and green for chlorine.

**2b. Feature Sets**

We evaluated eight representations of the iridium complexes to identify the most suitable set of features for training ML models to predict Ir phosphor properties. We compared eight feature sets: those based on similarity (i.e., Morgan, Dice), those based on graph descriptors (i.e., ligand-only revised autocorrelation, RACs[51], whole-complex RACs, and Coulomb-decay RACs[52], CD-RACs), and those based on electronic structure calculations (i.e., xTB, ωPBEh, and B3LYP).

For the similarity feature sets, we generated Morgan fingerprints[53,54], which have been used previously in machine learning chemistry applications[55-60], by one-hot encoding of groups of atoms in a structure. We computed these with a radius of three and 2,048 bits on the isolated **CN** and **NN** ligands to capture the presence and absence of chemical substructures. We also generated Dice similarity coefficients[53] of ligand Morgan fingerprints. In this approach, we compare the **CN** and **NN** ligands of each new iridium complex to all HLS **CN** or **NN** ligands through the Dice similarity metric (ESI Text S2 and Table S3). Dice similarity was selected after determining it outperformed the commonly employed Tanimoto similarity (ESI Table S4).

For the graph-based descriptor sets, we generated RACs[61,62] for both the isolated ligands and the full iridium complex structures (ESI Text S2). RACs are connectivity-based representations that have shown good performance for transition metal complex (TMC) property prediction.[41,61,63] For RACs, a TMC is represented as a molecular graph, with vertices for atoms and unweighted (i.e., no bond length or order information) edges for bonds. Each RAC feature is the sum of products or the sum of differences of heuristic atomic properties at depth $d$ on a TMC molecular graph, where $d$ indicates the number of edges separating the starting and ending atoms (ESI Text S2). The RACs include those that span over the entire complex as well as weighted averages over the equatorial ligands and axial ligands. We use the largest set of heuristic properties described in previous work, including both group number[64] and number of bonds[52], leading to a final RAC feature set that contains 196 features (ESI Table S5). For the ligand-only RAC feature set, we generated full-scope product RACs (i.e., all atoms are used as starting atoms) on isolated **CN** and **NN** ligands for each TMC. We concatenate individual feature vectors for the **CN** ligands and **NN** ligands (ESI Text S2). The ligand-only RAC feature set contains significantly fewer (i.e., only 70) features than the RAC feature set (ESI Table S6). We also generated Coulomb-decay RACs[65] on the iridium complex structures that were optimized with UFF (see *Dataset*). CD-RACs are a variant of RACs that also encode internuclear distance information between the atoms in the RAC feature (ESI Text S2). The CD-RAC feature set contains Coulomb-decay versions of the features in the RAC feature set but is of higher dimension (i.e., 222 features) due to the added information from the geometry (ESI Table S7).

Finally, we computed descriptors obtained from electronic structure theory. Specifically, we selected electronic properties of the isolated ligands due to the lower computational cost in comparison to whole-complex properties. These ligand-based decriptors include the HOMO and LUMO of each ligand type, the ionization potential (IP) and electron affinity (EA) of each ligand type, and the partial charges (i.e., Mulliken) of each of the metal-coordinating atoms (ESI Table S8). For the xTB feature set, we utilized a specially reparametrized[66,67] version of GFN1-xTB, a low-cost, semi-empirical tight binding method that has parameters for most elements in the periodic table.[68] We calculated ligand-only xTB features on UFF-optimized **CN** and **NN** ligands. We confirmed the suitability of ligand-only feature calculation due to the lower computational cost. The xTB features consist of contain electronic structure information on the **CN** and **NN** ligands of a phosphor, and are in some cases correlated (ESI Table S8 and Figure S2). For the B3LYP and ωPBEh DFT feature sets, we performed density functional theory (DFT) calculations on isolated **CN** and **NN** ligands using the B3LYP[69-71] or ωPBEh[72] exchange correlation functionals respectively (see *Methods*). Mulliken charges were used after they were found to outperform natural charges on the grouped split test set (ESI Table S9).

## 3. Results and Discussion.

### 3a. Features for ANN Models and Performance.

The representation for the phosphor is a crucial piece in determining which machine learning (i.e., ANN) models are most likely to both predict experimental properties accurately and generalize to unseen complexes. A model and feature set that performs well on one property may perform poorly on another. We thus trained models with each of the eight feature sets and the three target properties (Em$_{50/50}$, excited state lifetime, and emission spectral integral) and assessed their prediction performance on both a random split and grouped split. For the first random split, the Dice, Morgan, and xTB feature sets lead to the lowest model test set errors across all three target properties. Conversely, the B3LYP DFT and RAC feature sets lead to the largest model test set errors, while the remaining feature sets CD-RAC, ligand-only RAC, and ωPBEh DFT exhibit intermediate performance (Figure 2 and ESI Figures S3 and S4 and Tables S10-S13). The composition-based Dice and Morgan feature sets lead to the best predictions, judged on the basis of scaled MAEs of 0.03 to 0.05 for the three target properties (ESI Tables S11-S13). There is a substantial difference in performance: the percent difference in random split mean absolute error (MAE) between using the optimal feature set and the worst feature set for mean emission energy is 80% and a similar performance erosion is observed for the other two properties (i.e., 59% for spectral integral and 40% for phosphorescence lifetime).

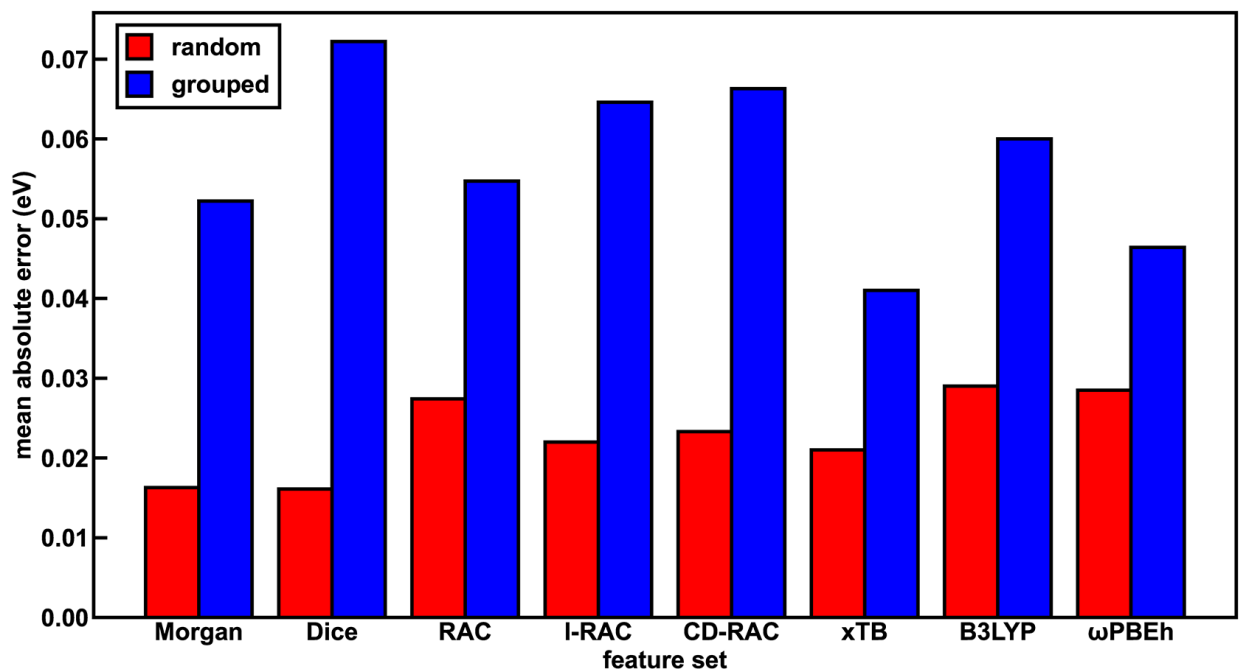

**Figure 2.** The test set performance of ANNs trained on different feature sets in predicting for Em$_{50/50}$ (MAE, in units of eV) for both random (red bars) and grouped splits (blue bars).

We rationalize the relative performance of each feature set in the random split assessment by considering what aspects of ligands and complexes the different feature sets capture. We attribute the predictive power of the Dice feature set to the fact that phosphorescent properties are very ligand-directed, and in the random split, each ligand is represented in both the train and test sets. We attribute the predictive power of the Morgan feature set in the random split to the identification of substructures in ligands that affect phosphor excited state properties by tuning energy levels and ligand rigidity. Thus, features encoding ligand similarity to previously observed ligands and substructures present in ligands lead to the best performance on the random split. On the other end of the spectrum, the random split ANNs trained on the RAC feature set (scaled MAE: 0.05 to 0.08) likely perform poorest because they encode significant metal-local information that does not vary across this set, since all complexes have an iridium center and an identical first coordination shell (ESI Tables S5 and S11-S13).

With regard to the ligand-only electronic structure feature sets, we surprisingly observe improved random split ANN performance with the xTB feature set relative to the two DFT feature sets. While xTB is expected to be faster than DFT for feature generation, electronic structure properties from xTB alone should not necessarily be more accurate. The $Em_{50/50}$ xTB model error is 30% lower than that of the corresponding B3LYP DFT model (i.e., 0.021 eV vs 0.029 eV). Nevertheless, most (i.e., eight of twelve) xTB features have high (>0.5) linear correlation with their B3LYP DFT and ωPBEh DFT counterparts, as determined by Pearson correlation coefficients (ESI Table S14). Thus, the GFN1-xTB method provides reliable electronic structure information from which our ANNs can generate accurate predictions, even if this electronic structure information is not of the same accuracy as that generated by DFT.

To assess the utility of these models for discovery of out-of-distribution complexes, we next used the feature sets to train new ANNs on a grouped split, where five ligands are present only in the test set and are consequently unseen by the ANNs during training (see Methods). We used the same grouped split across each property prediction task. We find that the test accuracy of the grouped split ANNs is worse than that of the corresponding random split ANNs for all features, due to the presence of unseen ligands in the grouped split test sets (ESI Tables S15-S17). However, this worsened performance is more significant for the spectral integral and $Em_{50/50}$ target properties. Overall, the MAE averaged over all feature sets is significantly worse for these two properties (157% or 164% worse on average for spectral integral and $Em_{50/50}$, respectively) than for phosphorescence lifetime (28% worse, ESI Table S18).

With the grouped split, the predictive power of the xTB feature set is further improved relative to the other feature sets (Figure 2 and ESI Figures S3 and S4). For $Em_{50/50}$ prediction, the xTB feature set improves from the third best feature set to the best feature set as a result of its scaled MAE increasing less than the random split best performers (i.e., 0.04 to 0.078 for xTB versus Dice 0.031 to 0.138, ESI Tables S18 and S19). In practice, this means that the $Em_{50/50}$ xTB MAE doubles from 0.021 eV to 0.041 eV, while the Dice MAE nearly quadruples from 0.016 eV to 0.072 eV. We attribute this particularly worsened performance of the Dice feature set to the reduction of the total number of features in the Dice feature set in the grouped split (ESI Table S3). The poor generalizability of the Dice feature set can also be attributed to the pseudo one-hot encoding of ligands via the similarity scores. The xTB features, in contrast, convey physical information that extrapolates beyond the ligands seen in the training data. We ultimately chose the xTB feature set for further analysis in evaluating hypothetical complexes, because the xTB feature set has favorable performance on the grouped split for all three properties, indicating that the ANNs using the xTB feature set generalize well.

To quantify ANN uncertainty in predictions for new phosphors outside of our initial training set, we use the latent space distance as a measure of how similar a new phosphor is to the phosphors used to train the model.[73] To confirm that this is a good measure of similarity that quantifies uncertainty for the current prediction task, we assessed the influence of latent space distance on test set prediction accuracy of the random split ANNs that use xTB inputs. Following prior work[73], we computed the average distance to ten nearest neighbors in the latent space formed by the training set and discarded predictions on any test set phosphor with an uncertainty quantification (UQ) metric exceeding the cutoff. A near monotonic decrease in average model error versus UQ cutoff suggests the possibility to control the error of predictions on new phosphors by discarding any prediction with a large UQ metric (Figure 3 and ESI Figures S5 and S6). Based on analysis of this UQ metric, we choose to avoid model prediction on novel complexes when the UQ is above the mean plus two standard deviations of the UQ metric evaluated over the random split test set complexes (see New Compound Exploration). Starting from a rescaled UQ metric where the most distant test complex is 1.0, this suggested cutoff corresponds to the largest value for the random split spectral integral ANN (i.e., 0.79) and somewhat smaller for the random split lifetime and $Em_{50/50}$ ANNs (i.e., 0.67 and 0.62).

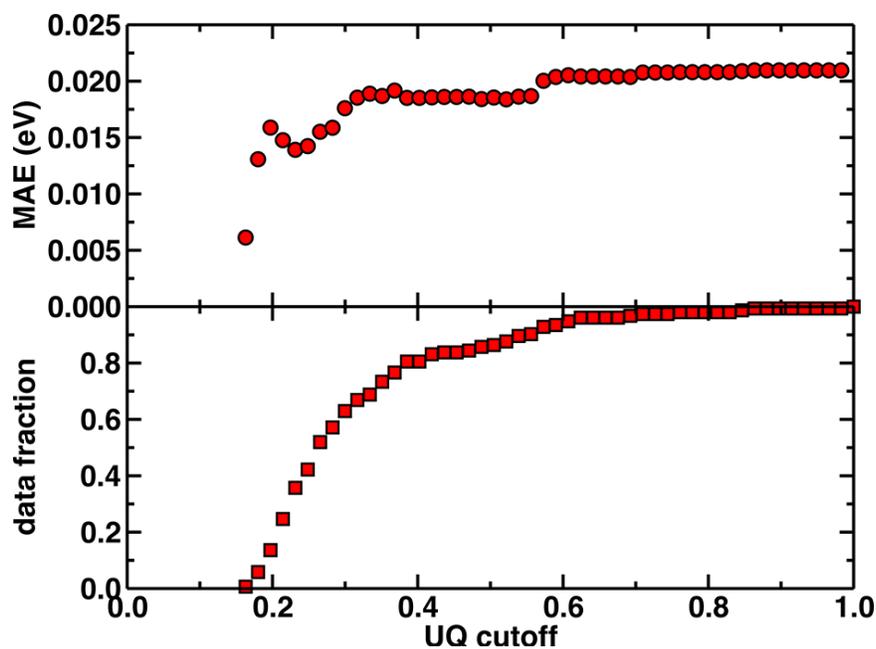

**Figure 3.** The uncertainty quantification (UQ) cutoff versus test set mean absolute error (in eV) of the random split ANN model trained on the xTB feature set and predicting $Em_{50/50}$. The data fraction is the number of test set complexes under the corresponding UQ cutoff, and the MAE is calculated on this subset of complexes. The UQ metric used is the average latent space distance to the ten nearest neighbors in the training set following the protocol introduced in Ref. [74]. The UQ metric is normalized such that the largest UQ metric is scaled to 1.

### 3b. Feature Importances and Trends

Given the high accuracy of the xTB-trained ANN models, we next sought to determine if simpler and more interpretable linear and random forest models trained on xTB features could attain similar accuracy. These models allow us to more transparently gain insight into which features most heavily influence phosphor property prediction. We trained random forest regression models that use xTB features to predict each of the three target properties using scikit-learn.[75] Using the same random split train/test partition, these random forest models have comparable performance to the ANNs, indicating that they can be analyzed to determine which features inform phosphor properties (ESI Table S20). Random forest models significantly outperform linear ridge regression models, which preclude their use for feature analysis but could have been anticipated from the low linear correlation of individual xTB features to target properties (ESI Table S20 and Figure S7). From random forest models, we extracted impurity-based feature importances to interpret the ML model predictions (Figure 4). We find that xTB features of the **CN** ligand are more important than those of the **NN** ligand in predicting Em$_{50/50}$ and lifetime. For both of these target properties, the sum of impurity-based importances of **CN** ligand features is approximately 50% larger than the corresponding sum for **NN** ligand features. The large role of the **CN** ligand in determining Em$_{50/50}$ can be explained by the partial localization of the phosphor complex HOMO on the **CN** ligand.[46] In contrast, xTB features of the **CN** and **NN** ligand are equally important in predicting the spectral integral. This indicates that when tuning Em$_{50/50}$ and lifetime, emphasis should be placed on selecting the **CN** ligand, whereas the **CN** and **NN** ligands play an equally large role in determining spectral integral.

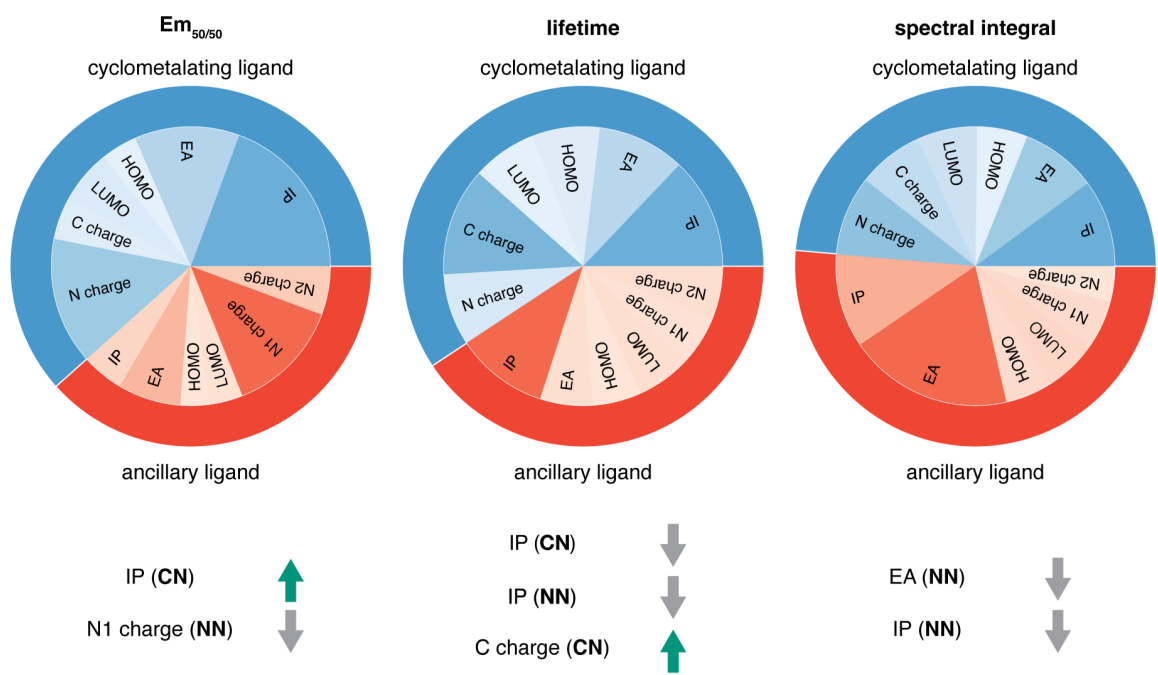

**Figure 4.** For each of the three target properties, the corresponding column indicates: (top) random forest feature importances of the xTB **CN** and **NN** features and (bottom) the correlation of the most important xTB features to the target property, where a green arrow indicates positive correlation and a gray arrow indicates negative correlation. For example, IP (**CN**) is positively correlated to Em$_{50/50}$, while N1 charge (**NN**) is negatively correlated to Em$_{50/50}$.

Focusing more on $Em_{50/50}$, we find that ionization potential (IP) and electron affinity (EA) are important for model predictions, as are the charges of metal-coordinating atoms (Figure 4). Specifically, the top three xTB features for predicting $Em_{50/50}$ are the IP of the **CN** ligand and two of the nitrogen-coordinating charges (i.e., N charge (**CN**) and N1 charge (**NN**)). This suggests that the information provided by the charge features allows the $Em_{50/50}$ ML model to make better predictions. The information provided by IP (**CN**) for the cyclometalating ligand also is important for model performance, and this conforms to prior observations that ligand energy levels affect emission energy.[76,77] We also emphasize that these three xTB features vary significantly over the experimental dataset. The IP (**CN**) varies by nearly 1.5 eV (i.e., from 7.56 eV to 9.03 eV), and the partial charges have a 0.1 a.u. range (i.e., N charge (**CN**) from -0.35 a.u. to -0.24 a.u. and N1 charge (**NN**) from -0.37 a.u. to -0.28 a.u) (Figure 5 and ESI Figure S8). Thus, tuning these three features in a coordinated fashion should enable tuning of Ir phosphor complex $Em_{50/50}$.

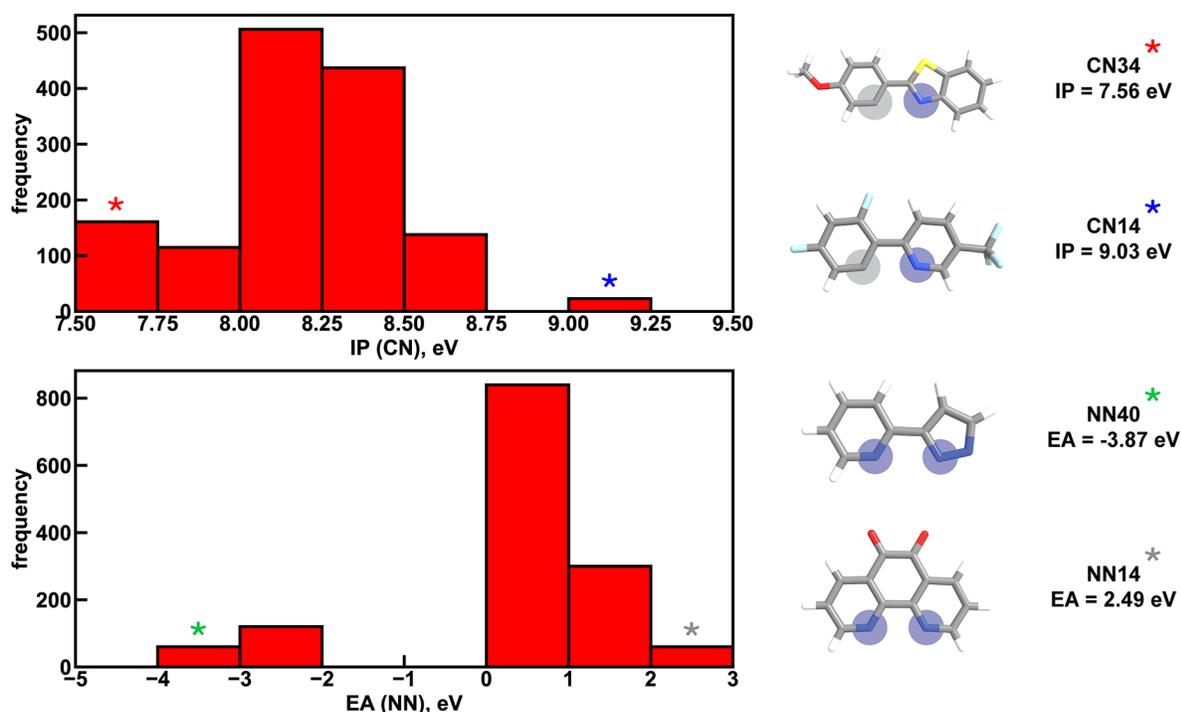

**Figure 5.** Distribution of two xTB features across the experimental dataset of 1,380 iridium phosphors. IP (**CN**) refers to the ionization potential of the **CN** ligand and EA (**NN**) refers to the electron affinity of the **NN** ligand. Asterisks correspond to ligands at the extreme ends of the distributions, shown on the right. Coordinated nitrogen (carbon) atoms are indicated with blue (gray) circles. Atoms are colored as follows: white for hydrogen, gray for carbon, blue for nitrogen, red for oxygen, light blue for fluorine, and yellow for sulfur.

As was observed from our global analysis, the most important xTB features are different for lifetime and spectral integral predictions (Figure 4). For predicting lifetime, IP features from both **CN** and **NN** ligands dominate, and the most important charge feature is the C charge of the **CN** ligand. For spectral integral, the top three features are EA (**NN**), IP (**NN**), and IP (**CN**), none of which are obtained from charges. The different feature importances for different target properties suggest some possibility of orthogonal design, wherein one phosphor property is tuned independently of the others. Nevertheless, given that ionization potential and electron affinity of the **CN** and **NN** ligands play a large role for all three target properties, altering coordinating atom charge tuning without significantly altering the IP/EA is likely the most direct way to target changes in Em$_{50/50}$ or lifetime without altering the spectral integral.

Considering the most important xTB features as determined by random forest analysis, we further identified specific compounds with extreme (i.e., high or low) experimental properties and compared how their xTB-computed features differed. For Em$_{50/50}$, high emission energy complexes typically have a high IP (**CN**), while low emission energy complexes typically have a low IP (**CN**). The N1 charge (**NN**) tends to be more positive for low emission energy complexes than for high emission energy ones. However, it is more challenging to identify which features are most important for long lifetime. In general, complexes with long lifetimes have a lower IP for both **CN** and **NN** ligands combined with a higher C charge (**CN**), but there are numerous exceptions. In the case of spectral integral, EA (**NN**) and IP (**NN**) are lower for bright complexes with high spectral integrals.

To further identify specific examples of phosphors in the original experimental dataset that demonstrate the trends, we examined pairs of iridium complexes that differ only in the identity of one type of ligand (neutral complexes are generated when **NN40**, **NN41**, or **NN42** is the ancillary ligand; ESI Text S1). One such pair is [Ir(**CN67**)$_2$(**NN41**)]$^0$ and [Ir(**CN95**)$_2$(**NN41**)]$^0$ (Figure 6). The former complex has an IP (**CN**) of 8.67 eV due to the electron-withdrawing fluorine groups on the cyclometalating ligand, while the latter complex has an IP (**CN**) of 7.67 eV. These values are on opposite ends of the IP (**CN**) distribution and contribute to Em$_{50/50}$ values on opposite ends of Em$_{50/50}$ distribution, 2.45 eV and 2.12 eV, respectively (Figure 5 and ESI Figure S1). The remaining five **CN** features for these two phosphors do not differ greatly from one another, underscoring the overriding effect of IP (**CN**). Similarly, increasing EA (**CN**) and EA (**NN**) can have a large effect on lifetime and spectral integral respectively (ESI Figures S9 and S10). These examples illustrate how differences in xTB features caused by ligand substitution correlate to shifts in phosphor properties.

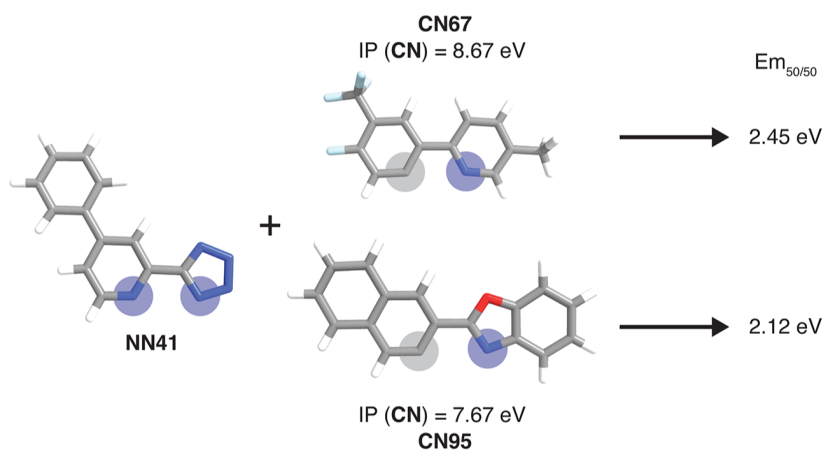

**Figure 6.** Example of a pair of complexes where the substitution of the **CN** ligand leads to a large Em$_{50/50}$ property change. Coordinated nitrogen and carbon atoms are indicated with blue and gray circles, respectively. The relevant xTB features for the substituted ligands are shown. Atoms are colored as follows: white for hydrogen, gray for carbon, blue for nitrogen, red for oxygen, and light blue for fluorine.

To determine how ligand selection can allow for independent tuning, we consider the four complexes [Ir(**CN101**)$_2$(**NN2**)]$^+$, [Ir(**CN101**)$_2$(**NN20**)]$^+$, [Ir(**CN105**)$_2$(**NN2**)]$^+$, and [Ir(**CN105**)$_2$(**NN20**)]$^+$ that each differ by a single ligand. Changing the cyclometalating ligand from **CN101** to **CN105** leads to an increase in Em$_{50/50}$ while having a small effect on phosphorescence lifetime, while changing the ancillary ligand from **NN2** to **NN20** leads to an increase in phosphorescence lifetime while having a small effect on Em$_{50/50}$ (Figure 7). The increase in Em$_{50/50}$ when swapping **CN101** for **CN105** and the increase in lifetime when swapping **NN2** for **NN20** follows our observed trends of IP (**CN**) correlating positively to Em$_{50/50}$ and IP (**NN**) correlating negatively to lifetime. Furthermore, the small change in Em$_{50/50}$ when changing from **NN2** to **NN20** can be rationalized by the similar N1 charge (**NN**) between the two ancillary ligands. This example demonstrates how phosphor properties can be tuned orthogonally as guided by xTB features.

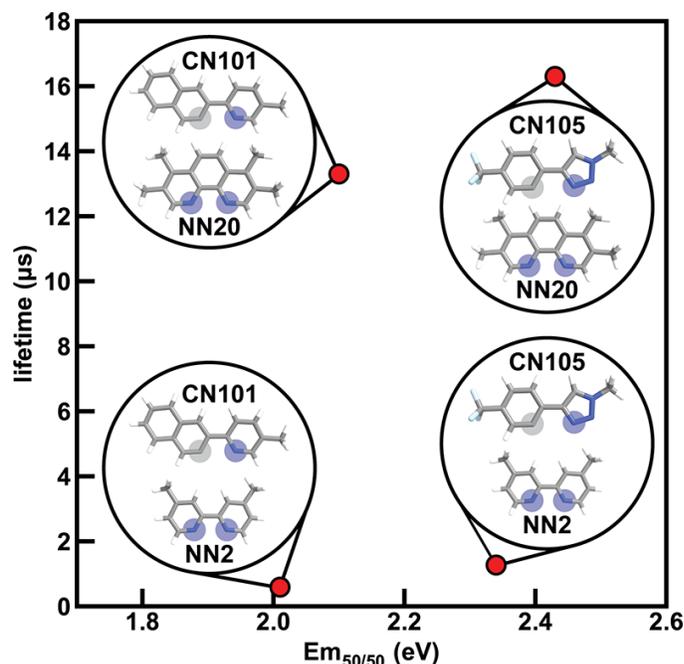

**Figure 7.** Four iridium phosphor complexes and the effect of substituting out the **CN** or **NN** ligand on Em$_{50/50}$ and lifetime indicated in the graph with structures shown in inset. Coordinated nitrogen (carbon) atoms are indicated with blue (gray) circles. Atoms are colored as follows: white for hydrogen, gray for carbon, blue for nitrogen, and light blue for fluorine.

### 3c. New Compound Exploration

We applied the three random split ANNs that use xTB features as inputs to screen hypothetical iridium complexes generated from CSD ligands (see Methods and ESI Text S1). We only considered hypothetical complexes under the UQ cutoff for all three ANNs (see Uncertainty Quantification). From our CSD screen, we identified 153 unique out-of-HLS **CN** ligands and 269 unique out-of-HLS **NN** ligands were identified and added to the 60 **CN** and 23 **NN** ligands from the HLS set. Combining these sets led to 60,816 hypothetical complexes with at least one out-of-HLS ligand, which reduced to 3,598 hypothetical complexes after applying the UQ cutoff. This corresponds to inclusion of 70 unique out-of-HLS **CN** ligands and 42 unique out-of-HLS **NN** ligands in combination with each other or with HLS **CN** and **NN** ligands.

For this set of curated hypothetical complexes, we evaluated which ligands are present in the complexes with the highest and lowest ANN-predicted properties (ESI Figure S11). We find that specific ancillary ligands tend to be well-represented in complexes with extreme properties, indicating that phosphor properties are tuned by these ancillary ligands (ESI Table S21). For example, the ligand that appears most often in hypothetical complexes with high predicted lifetime is the ancillary ligand from the CSD structure with refcode RASGAV. This conjugated ligand has a relatively low IP (**NN**) of 7.79 eV, which contributes to a longer lifetime following the previously identified trend (Figure 4 and 8). Indeed, the other ancillary ligands that are well-represented in hypothetical complexes with extreme predicted lifetimes (**NN** ligands from complexes with refcodes FEQSEB, MIMYEO, TOTPAW, OVALEE, and MAXWIS) also follow the trend of low IP (**NN**) correlating to long lifetime (ESI Table S21). We also note clear xTB feature trends in predictions for spectral integral and $Em_{50/50}$. The low IP (**NN**) of the ancillary ligand from RASGAV leads to a hypothetical complex with one of the highest predicted spectral integrals (Figure 4 and 8 and ESI Table S22). With regard to $Em_{50/50}$, the fluorinated cyclometalating ligand from the CSD structure with refcode RADTEZ has a high ionization potential (9.24 eV). The high IP (**CN**) feature appears to contribute to a high emission energy, as the RADTEZ **CN** ligand is present in the three highest predicted Em$_{50/50}$ hypothetical complexes (Figure 4 and Figure 8 and ESI Table S22). On the other hand, the ancillary ligands LEZJAD **NN** and TUZHEE **NN** have high N1 charge (**NN**) features, leading to their presence in the three lowest predicted Em$_{50/50}$ hypothetical complexes (Figure 4 and Figure 8 and ESI Table S22). Thus, we find that many ligands that lead to extreme hypothetical phosphor predicted properties follow our identified xTB feature trends from the experimental data. This lends interpretability to our model predictions and indicates that these predictions are derived from the electronic structure properties of the ligands.

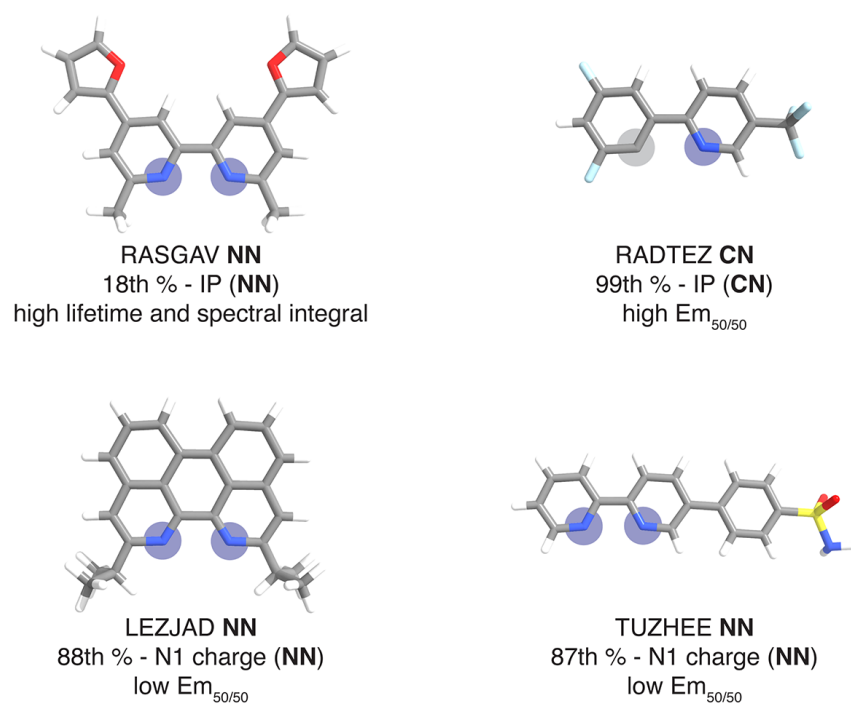

**Figure 8.** Ligands mined from the CSD that lead to very high or very low phosphor properties predicted by the ANNs along with their percentile rank of the relevant property in the context of the experimental complexes. Coordinated nitrogen and carbon atoms are indicated with blue and gray circles respectively. Atoms are colored as follows: white for hydrogen, gray for carbon, blue for nitrogen, red for oxygen, light blue for fluorine, and yellow for sulfur.

To further validate random split ANN predictions, we obtained TDDFT excited state energy and lifetime predictions and compared them to ANN predictions over complexes in both the experimental dataset and the hypothetical dataset. We found that using singlet geometries instead of triplet geometries as inputs to TDDFT leads to better agreement with experiment, although geometries do not differ greatly from a RMSD perspective (ESI Table S23 and Figure S12). Emission energies calculated with B3LYP were found to correlate with experiment better than those calculated with the range-separated hybrid functional CAM-B3LYP, motivating our use of B3LYP for TDDFT (ESI Figure S13).

Over a group of 26 representative test set complexes from the experimental dataset, we find that TDDFT overestimates experimental emission energy by 0.3 eV on average, and further find that TDDFT predictions correlate with experiment less well than the $Em_{50/50}$ ANN predictions (Figure 9 and ESI Tables S24 and S25). These results show that the $Em_{50/50}$ ANN achieves excellent performance. Even after applying a rigid downward shift to TDDFT energy predictions, they exhibit a larger spread around the experimental values than the predictions of our $Em_{50/50}$ ANN. Over the same 26 complexes, TDDFT lifetime predictions trend with experiment and ANN predictions; however, unlike the case of $Em_{50/50}$, TDDFT predictions outperform our lifetime ANN for complexes with long lifetimes (ESI Figure S14). This shortcoming of the lifetime ANN can be rationalized by the lower amount of phosphors with long lifetimes in the experimental dataset used for model training (ESI Figure S1 and Table S26). Although we do not have ground truth experimental data on our hypothetical set, we can still use TDDFT predictions to validate our ANN models. Over 21 representative hypothetical complexes, TDDFT energy predictions are on average 0.14 eV above $Em_{50/50}$ ANN predictions, and TDDFT lifetime predictions trend with lifetime ANN predictions (ESI Figures S15-S17 and Tables S27 and S28). These results indicate ANN predictions over the hypothetical set of complexes are reliable, although they may underestimate the lifetime of phosphors with long lifetimes. Thus, the ANNs are a trustworthy tool for the identification of hypothetical complexes with desired excited state properties.

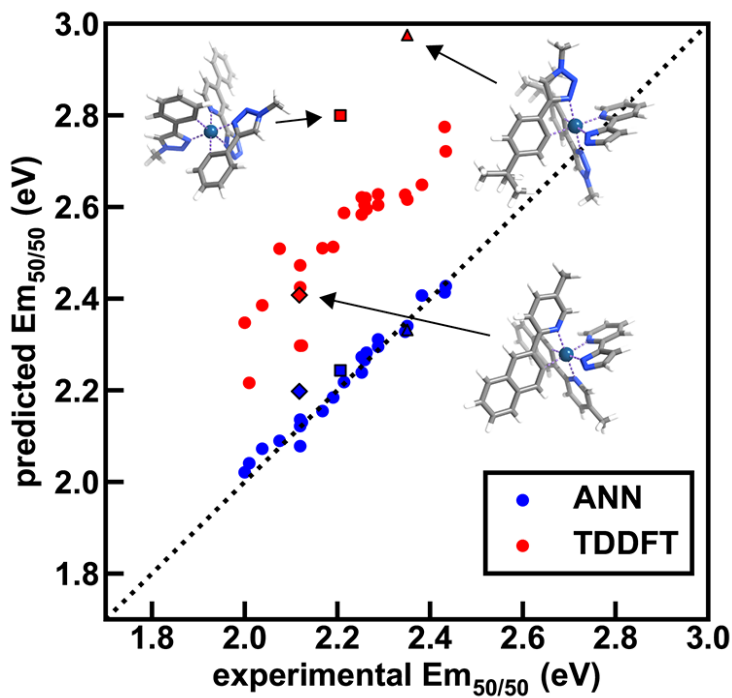

**Figure 9.** Comparison of random split ANN and TDDFT $Em_{50/50}$ prediction to experiment (in eV) across 26 test set iridium complexes in the experimental dataset. These complexes were chosen to span the range of emission energies and lifetimes of the full set. TDDFT was carried out on optimized singlet geometries using the B3LYP functional. Three high error complexes ([Ir(**CN101**)$_2$(**NN40**)]$^0$, [Ir(**CN107**)$_2$(**NN41**)]$^0$, and [Ir(**CN109**)$_2$(**NN40**)]$^0$) are shown as insets, and their predicted and experimental $Em_{50/50}$ values are shown with black borders and unique shapes (diamond, square, and triangle, respectively). In the insets, atoms are colored as follows: white for hydrogen, gray for carbon, blue for nitrogen, and dark blue for iridium. The dotted line is included as a reference and corresponds to perfect agreement between prediction and experiment.

**4. Conclusions.**

While *ab initio* methods like TDDFT are useful tools for studying iridium phosphor excited states, they are both computation-intensive and can also have eroded accuracy, motivating the use of machine learning. Using experimental data on 1,380 iridium phosphors, we trained ANNs to predict three experimental properties: $Em_{50/50}$, excited state lifetime, and emission spectral integral.

We found that features calculated with xTB led to the best overall performance across the three properties on out-of-sample complexes, outperforming the standard Morgan fingerprint features and features based on RACs. We then used random forest regression models to determine which xTB features most influence phosphor properties, deriving chemical insight. We found that high cyclometalating ligand ionization potential is indicative of high $Em_{50/50}$, while high ancillary ligand ionization potential correlates to low lifetime and low spectral integral. These observations illustrate how phosphor properties can be altered through judicious ligand selection.

We next demonstrated how our ANNs can be applied to chemical exploration by considering hypothetical iridium phosphors derived from ligands found in the CSD. We identified cyclometalating and ancillary ligands that lead to edge-of-distribution properties, such as an ancillary ligand predicted to result in both high lifetime and spectral integral phosphors. We confirmed the validity of these predictions by comparing to TDDFT, showing that for $Em_{50/50}$ the ANN significantly outperforms TDDFT, while for lifetime the corresponding ANN performs well only in regimes of sufficient training data. The ANN models for iridium phosphor property prediction that we present here are promising tools for chemical screening and the acceleration of chemical discovery, as they can be used to quickly evaluate thousands of hypothetical iridium phosphors to identify promising candidates for follow-up synthesis.

## 5. Computational Details.

### 5a. Feature and Structure Generation

We generated feature sets to represent the 1,380 Ir phosphor complexes as inputs to ML models (see *Feature Sets*). We generated all ligands using the draw tool in Avogadro v1.1.2[47,48] and subsequently optimized them with UFF.[78] We used molSimplify v1.6.0 for the generation of RAC feature sets on either ligands or complexes.[49,50] The xTB features were generated using xTB 6.4.0[68], and DFT-based features were generated using the B3LYP[69-71] or ωPBEh[72] functional with the LACVP* basis set implemented in the TeraChem v1.9-2018.11-dev[79,80] program. For the generation of the Dice feature set, we used RDKit 2021.9.2[81] both for Morgan fingerprints and the Dice similarity coefficients evaluated on Morgan fingerprints. For electronic structure descriptors, ligand geometries were geometry minimized with neutral charge and singlet spin multiplicity using DFT in TeraChem. Geometry optimizations used the L-BFGS algorithm in translation rotation internal coordinates (TRIC)[82] as implemented in TeraChem to the default tolerances of $4.5 \times 10^{-4}$ hartree/bohr for the maximum gradient and $1 \times 10^{-6}$ hartree for the change in energy between steps. We then performed single point energy calculations on the optimized neutral ligand geometries at two different charges: +1 and -1 (ESI Text S1). We used the LACVP* basis set, which for the HLS ligands corresponds to the LANL2DZ[83] effective core potential for Br and the 6-31G* basis set for all remaining elements. We calculated all non-singlet states with an unrestricted formalism and singlet states with a restricted formalism. Level shifting of 0.25 Ha was employed on both virtual and occupied orbitals to facilitate self-consistent field convergence. We used the hybrid DIIS[84]/A-DIIS[85] scheme for the self-consistent field procedure. We used TeraChem dynamic precision and a grid with approximately 3,000 points per atom. Like the xTB feature set, DFT-generated features encode electronic structure information (ESI Table S8).

**5b. ML Models**

We trained multiple artificial neural networks (ANNs) with for each of the seven feature sets to predict three target properties: $Em_{50/50}$, excited state lifetime, and emission spectral integral. For all random split ANNs, we used a random 70%/15%/15% train/validation/test split of the 1,380 complexes from the prior study[46] and refer to this hereafter as the random split. To assess the generalizability of our best-performing ANNs, we also carried out grouped splits where we excluded from the training and validation data any complex containing a ligand from a select subset of **CN** and **NN** ligands. For the excluded ligands, we selected **CN21**, **CN103**, **CN104**, **NN20**, and **NN43** after determining these to be the most dissimilar HLS ligands relative to the other HLS ligands as measured through Dice similarities of Morgan fingerprints (ESI Tables S1-S2 and S30). If one of the 1,380 complexes contains one or more of these ligands, it is held out from the test set. We pre-processed features by normalizing each feature to a zero mean and unit variance over the train and validation data and removed any invariant features (ESI Text S2). For ANNs predicting for lifetime and $Em_{50/50}$, we excluded 356 complexes with low luminescent intensity (i.e. spectral integral less than $1\times10^5$ counts) from ANN training and performance evaluation due to the greater noise in lifetime and $Em_{50/50}$ measurements for dim Ir phosphors.

We built ANNs with Keras 2.4.3 with TensorFlow 2.3.0 as the backend.[86,87] Hyperparameters for each ANN were chosen using Hyperopt[88] with 200 evaluations, as judged by the mean absolute error of the model on the validation data. The built-in tree of Parzen estimator[89] algorithm in Hyperopt was used to select model hyperparameters. We used these chosen hyperparameters to train the final model on the combined train and validation data and evaluated performance on the test set (ESI Table S29). All ANN models were trained with the AMSGrad variant[90] of the Adam optimizer[91] up to 2000 epochs. Dropout[92], batch normalization[93], and early stopping[94] were applied to avoid over-fitting. The patience for early stopping was 100. We enforced a floor of zero for all predictions since negative predictions for $Em_{50/50}$, lifetime, or spectral integral are unphysical. All machine learning models have been deposited online in a Zenodo repository.[47]

**5c. Out-of-distribution Complexes**

We identified hypothetical out-of-distribution iridium complexes which we enumerated combinatorially using **CN** and **NN** ligands not in the HLS. We selected these ligands by screening the CSD v5.42 + 2 updates, released in November 2020, for iridium complexes with two **CN** ligands and one **NN** ligand by specifying the first coordination sphere around iridium in a ConQuest 2021.1.0 search. Complexes selected by the screening were then examined by hand, and those that were not fit for analysis were eliminated (ESI Table S31). The molSimplify code was used to identify unique ligands from the remaining complexes on the basis of their atom-weighted molecular graph determinants[95], and we used any ligands not already in the HLS in combination with the HLS to generate new hypothetical $[Ir(CN)_2(NN)]^+$ complexes (ESI Text S1).

**5d. TDDFT Calculations**

For *ab initio* validation of predictions using TDDFT, iridium phosphors were first geometry optimized, and TDDFT was then run on the optimized geometries using the ORCA 5.0.1[96] program. All calculations employed a C-PCM solvation correction[97] to mimic DMSO. Singlet geometry optimization was carried out using the B3LYP[69-71] functional and the def2-TZVP[98] basis set with D4 dispersion correction[99] on structures generated by molSimplify. For TDDFT, the B3LYP functional and the Zero-Order Regular Approximation (ZORA)[100] were used. The SARC-ZORA-TZVP[101] basis set was used for iridium and the ZORA-def2-TZVP basis set was used for all other elements along with the SARC/J auxiliary basis set. The TDDFT calculation included 25 roots, quasi-degenerate perturbation theory spin-orbit coupling[102] was enabled, and the Tamm-Dancoff approximation was disabled.

Due to relativistic SOC caused by iridium, the T1 manifold is split into three sublevels (zero-field splitting). For the calculation of *ab initio* energy, the energies of these three lowest triplet sublevels from the TDDFT calculation were averaged for each complex. For *ab initio* lifetime, radiative rate and radiative lifetime were calculated as in prior work[27-29,31,103] using output from TDDFT calculations. The radiative rate $k_i$ from a triplet sublevel $i$ is given by:

$$k_i = \frac{1}{\tau_i} = \frac{4}{3t_0}\alpha_0^3(\Delta E_i)^3 \Sigma_{\alpha\in\{x,y,z\}}|M_\alpha^i|^2 \qquad (1)$$

where $\tau_i$ is the radiative lifetime of sublevel $i$, $t_0 = \frac{(4\pi\epsilon_0)^2\hbar^3}{m_e e^4}$, $\alpha_0$ is the fine structure constant, $\Delta E_i$ is the excitation energy in atomic units from the ground state to the sublevel $i$, and $M_\alpha^i$ is the $\alpha$-axis projection of the transition dipole moment in atomic units between the ground state and the sublevel $i$.

The overall radiative lifetime from the three triplet sublevels is calculated as a Boltzmann average of radiative rates that depends on the energy differences between triplet sublevels.

$$\tau_{av} = \frac{1}{k_{av}} = \left(\frac{1 + e^{-(\Delta E_{1,2}/k_B T)} + e^{-(\Delta E_{1,3}/k_B T)}}{k_1 + k_2 e^{-(\Delta E_{1,2}/k_B T)} + k_3 e^{-(\Delta E_{1,3}/k_B T)}}\right) \qquad (2)$$

$\Delta E_{1,2}$ is the energy difference between sublevels 1 and 2, and $\Delta E_{1,3}$ is the energy difference between sublevels 1 and 3. T = 300K was used. This equation for lifetime does not take into account nonradiative decay, which can be significant in some cases. In order to account for the DMSO solvent, the *ab initio* lifetime was divided by the square of the refractive index of DMSO according to the Strickler-Berg relationship.[104]

ASSOCIATED CONTENT

**Electronic supplementary information**. Information about DFT calculations; information about the CSD iridium phosphor search; information about the feature sets; histograms of the target properties in the experimental dataset; correlation of xTB features with themselves, with target properties, and with DFT features; MAE of ANNs trained on different feature sets in predicting for lifetime and spectral_integral; the effect of UQ cutoff on model accuracy in predicting for lifetime and spectral integral; distribution of xTB features; effect of ligand substitution on lifetime and spectral integral; ligands present in complexes with extreme predicted properties; comparison of singlet and triplet geometries of phosphors; CAM-B3LYP TDDFT energy predictions; comparison of experiment, ANN predictions, and TDDFT predictions for $Em_{50/50}$ and lifetime; confusion matrices for a 2 μs lifetime cutoff; structures of **CN** and **NN** ligands from the experimental dataset; information about the features in each feature set; comparison between different charge schemes and fingerprint similarity metrics; the ranking by MAE of different feature sets on the random and grouped split; performance of different feature sets on the random and grouped split; the change in model performance from random to grouped split for different feature sets; comparison between different ML models; correlation coefficients between experiment, ANN predictions, and TDDFT predictions; list of complexes used for TDDFT benchmarking; ANN predictions for complexes with long experimental lifetime; hyperparameters of the best-performing ANNs; most dissimilar HLS ligands as determined by Dice similarity; attrition of CSD complexes (PDF).

XYZ files of the **CN** and **NN** ligands from the experimental dataset; XYZ files of ligands mined from the CSD; xTB features of experimental and hypothetical phosphors; ANN-predicted values for the hypothetical phosphors; train/validation/test splits for the random and grouped splits; example Python scripts for featurization, ANN training, and ANN application (ZIP).

## AUTHOR INFORMATION

**Corresponding Author**

*email:hjkulik@mit.edu

## DATA AVAILABILITY

The datasets supporting this article have been uploaded as part of the ESI.

## AUTHOR CONTRIBUTIONS

Gianmarco Terrones: data curation, ML training, conceptualization, writing – original draft preparation, visualization; Chenru Duan: ML training, writing – reviewing and editing; Aditya Nandy: data curation, writing – reviewing and editing; Heather J. Kulik: writing – reviewing and editing, supervision, conceptualization.

## CONFLICTS OF INTEREST

The authors declare no competing financial interest.

ACKNOWLEDGMENT

The authors acknowledge primary support for this work from the Office of Naval Research under grant numbers N00014-18-1-2434 and N00014-20-1-2150. Support for machine learning feature development was also provided by DARPA under grant number D18AP00039. G.T. was partially supported by an Alfred P. Sloan Foundation Scholarship (Grant Number G-2020-14067). A.N. was partially supported by the National Science Foundation Graduate Research Fellowship Program (Grant Number #1122374). This work was carried out in part using computational resources from the San Diego Supercomputer Cluster (SDSC), and in part using computational resources from the Extreme Science and Engineering Discovery Environment (XSEDE) which is supported by National Science Foundation grant number ACI-1548562. The authors acknowledge Adam H. Steeves for providing a critical reading of the manuscript.

**Contents**

**Text S1.** Additional details of calculations and CSD search.

*Additional DFT calculation details*

As mentioned in the main text, we conducted single point energy calculations on the optimized neutral ligand geometries at two different charges: +1 and -1. In combination with information from the neutral geometry optimization, this allowed us to calculate vertical IP and EA values analogous to those generated by GFN1-xTB.[1,2] The remaining information (HOMO, LUMO, and Mulliken charges of coordinating atoms) was extracted solely from the neutral geometry optimization calculation for a given ligand. We specified ligands to be in a singlet state when neutral and in a doublet state otherwise.

*Description of NN40, NN41, and NN42*

These three ancillary ligands contain tetrazole or pyrazole moieties that deprotonate upon metal coordination.[3] Consequently, we generated these ligands without a hydrogen on the pyrazole/tetrazole nitrogen coordinating atom. As a result, while all other HLS ligands are neutral, these ligands have a charge of -1. We specified this charge for GFN1-xTB calculations, and also adjusted our DFT workflow to account for it as follows: We performed geometry optimization on the **NN40**, **NN41**, and **NN42** ligand geometries generated with Avogadro while specifying a charge of -1. We then conducted single point energy calculations on the optimized anionic ligand geometries at two different charges: neutral and -2. We specified these ligands to be in a singlet state when at -1 charge and in a doublet state otherwise.

Iridium complexes with these ancillary ligands likely form neutral complexes $[Ir(\mathbf{CN})_2(\mathbf{NN})]^0$, rather than the +1 charge complexes $[Ir(\mathbf{CN})_2(\mathbf{NN})]^+$.

*Additional CSD details*

In a ConQuest search, we searched for structures containing an iridium atom bonded to two carbon atoms and four nitrogen atoms. We set bond types to "Any" and the cyclicity of the carbon and nitrogen atoms to "Cyclic." We searched with 3D coordinates determined and an R factor <= 0.05. Complexes in the hitlist were exported as mol2 files with the "Export largest molecule only" and "One file per entry" options, where the former option removes solvent molecules. From 700 hits, those with multiple iridium atoms and no iridium atoms were removed from consideration, as were hits that still had solvent or counterions, hits that did not have three bidentate ligands, duplicate hits as determined by CSD refcodes, and hits that had a **CC** ligand (Table S31).

Classification of CSD ligands as **CN** or **NN** was accomplished by analyzing the exported mol2 files using molSimplify, which identifies coordinating atoms in addition to identifying ligands. Hydrogen atoms were added to coordinating carbons of CSD **CN** ligands using molSimplify. The presence or absence of a CSD ligand in the HLS was determined through atom-weighted molecular graph determinants, computed using molSimplify. CSD ligands absent from the HLS will be referred to as out-of-HLS CSD ligands. Atom-weighted molecular graph determinants were also used to ensure each out-of-HLS CSD ligand structure was only considered once in hypothetical complex enumeration, since multiple hit complexes might have **CN** or **NN** ligands in common. Hypothetical complexes either had two HLS **CN** ligands and an out-of-HLS CSD **NN** ligand, two out-of-HLS CSD **CN** ligands and an HLS **NN** ligand, or two out-of-HLS CSD **CN** ligands and an out-of-HLS CSD **NN** ligand. The two **CN** ligands were always the same in any given complex. From the final complexes in Table S31, 153 unique out-of-HLS **CN** ligands and 269 unique out-of-HLS **NN** ligands were identified. Consequently, 60,816 hypothetical complexes were considered. Application of uncertainty quantification cutoffs left 70 unique out-of-HLS **CN** ligands and 42 unique out-of-HLS **NN** ligands spread out over 3,598 hypothetical complexes, which we analyzed with our ANNs.

**Text S2.** Extended description of feature sets.
*Explanation of feature notation.*

For a given iridium complex, features in the ligand-only RAC set, the xTB set, the B3LYP DFT set, the ωPBEh DFT set, and the Dice set require only the molecular geometry of the **CN** and **NN** ligand of the complex.

For RAC-style feature sets, the atomic properties considered were topology, identity, electronegativity, covalent radius, nuclear charge, group number, and number of bonds by the octet rule.[4] T is topology, I is identity, chi is electronegativity, S is covalent radius, Z is nuclear charge, Gval is group number, and NumB is number of bonds. For the RAC and CD-RAC feature sets, mc indicates metal-centered, lc indicates ligand-centered, D indicates difference (and its absence indicates product), and depth is indicated by the number in the feature name. The terms all, ax, and eq refer to the extent of a RAC, i.e. whether it spans over the whole complex or only over the axial or equatorial ligands. The four miscellaneous features that describe charge or denticity of a ligand were all removed. For the ligand-only RAC feature set, atom-wise properties, depth, and ligand type are indicated by the feature name.

For the electronic structure feature sets, IP stands for ionization potential and EA stands for electron affinity. The first coordinating nitrogen of the **NN** ligand (N1) is chosen such that the number of nitrogen atoms in its ring is less than or equal to the number of nitrogen atoms in the ring of the second coordinating nitrogen.

*Invariant features*

Invariant features, i.e. features that are the same across all complexes in the experimental training data, were removed during pre-processing. Only the Morgan, RAC, and CD-RAC feature sets had invariant features.

*Morgan and Dice feature sets*

A Morgan fingerprint indicates the presence of substructures in a molecule by hashing any given substructure into a X-bit integer, and effectively storing these integers as the indices of bits set to 1 in a $2^X$-size bitset.[5] The Morgan feature set initially contains 4,096 features, with 2,048 bits (X=11) allocated to both the **CN** ligand and **NN** ligand; however, over 75% of these features are invariant over the training data and are dropped from the set. The Dice coefficient between two Morgan fingerprints A and B is defined as $\frac{2c}{a+b}$, where a is the number of bits set to 1 in A, b is the number of bits set to 1 in B, and c is the number of bits set to 1 in both A and B[6]. We also considered the popular Tanimoto[6] coefficients ($\frac{c}{a+b-c}$) but found these to hold less predictive power than Dice coefficients in the present application.

*Revised autocorrelation functions details*

RAC features can be full-scope, metal-centered, or ligand-centered; these distinctions indicate the positions of the starting atoms used to generate the RAC feature. For a full-scope RAC feature, every atom in the complex can be used as the starting atom. For a metal-centered RAC feature, the metal center serves as the starting atom. For a ligand-centered RAC feature, the coordinating atoms on the ligands serve as starting atoms.

Mathematically, autocorrelations are defined as

$$P_d = \Sigma_i \Sigma_j P_i P_j \delta(d_{ij}, d)$$

(1)

$$P'_d = \Sigma_i \; \Sigma_j \; (P_i - P_j)\delta(d_{ij}, d)$$

(2)

where $P_d$ is the graph autocorrelation for property $P$ at depth $d$, $P'_d$ is the analogous difference graph autocorrelation, $\delta$ is the Dirac delta function, and $d_{ij}$ is the bond-wise path distance between atoms $i$ and $j$.

*RAC-style feature sets details*

For the RAC feature set, we allowed $d$ to range from zero – corresponding to the correlation of an atom property to itself – to three. The RAC feature set consists of 196 features.

For the ligand-only RAC feature set, we allowed $d$ to range from zero to four. We used a larger maximum depth than that of the RAC feature set due to the comparatively low number of features in the ligand-only case. This feature set was motivated by our anticipation that separate treatment of **CN** and **NN** ligands would improve predictive power. Furthermore, the absence of metal-centered information in this feature set is acceptable because it was anticipated that the metal-centered RACs in the RAC feature set would be uninformative - all complexes in this study have an iridium metal center and an identical first coordination sphere of two carbon atoms and four nitrogen atoms. The ligand-only RAC feature set consists of 70 features.

For the CD-RAC feature set, we allowed $d$ to range from zero to three. Both the RAC feature set and the ligand-only RAC feature sets can be generated from structures that are not geometry optimized. This is because RACs are connectivity-dependent features but do not depend on geometry information such as bond lengths. In contrast, CD-RACs are affected by geometry optimization. CD-RACs were calculated on molSimplify-generated structures optimized with UFF. This feature set was motivated by the impact of bond distances on Ir phosphor properties like quantum yield.[7] The CD-RAC feature set consists of 222 features.

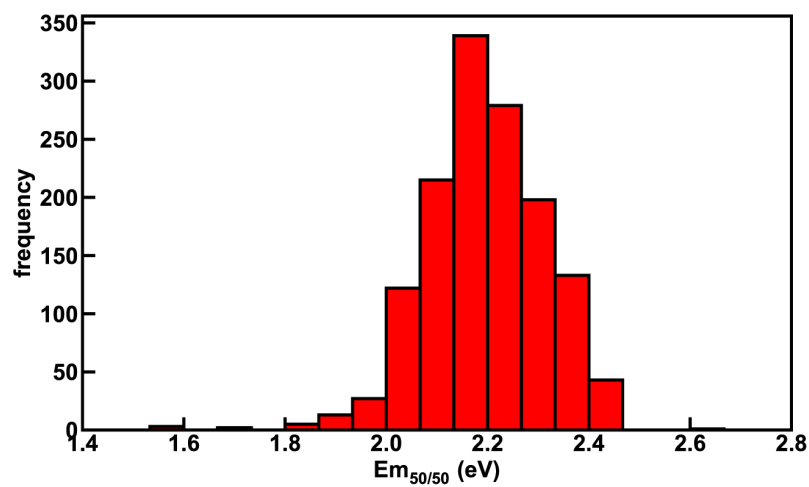

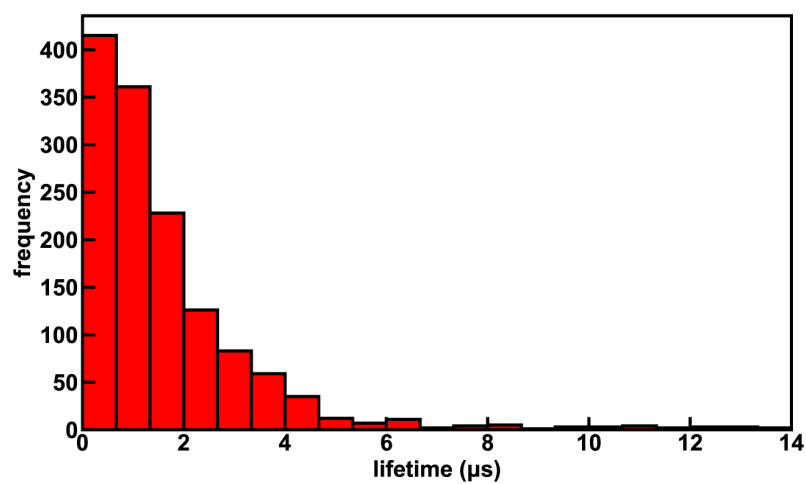

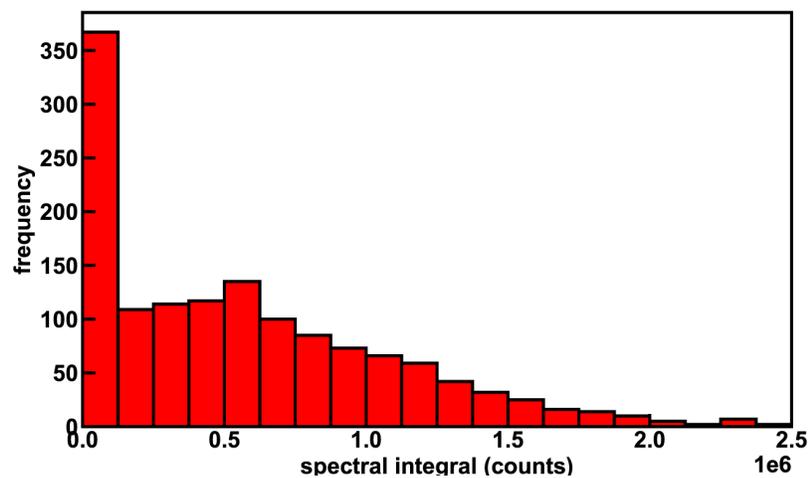

**Figure S1.** Histograms of the three target properties across the 1,380 complexes reported in the experimental study of DiLuzio *et al.*[3], excluding the baseline *solvato* complexes that contain a DMSO ligand. The range of phosphorescence lifetimes is restricted to omit eleven outliers with long lifetimes ranging from 14 to 75 μs. Complexes with low spectral integral (less than $1 \times 10^5$ photon counts) are considered dim in this work and are excluded from further $Em_{50/50}$ and lifetime analysis (see Methods).

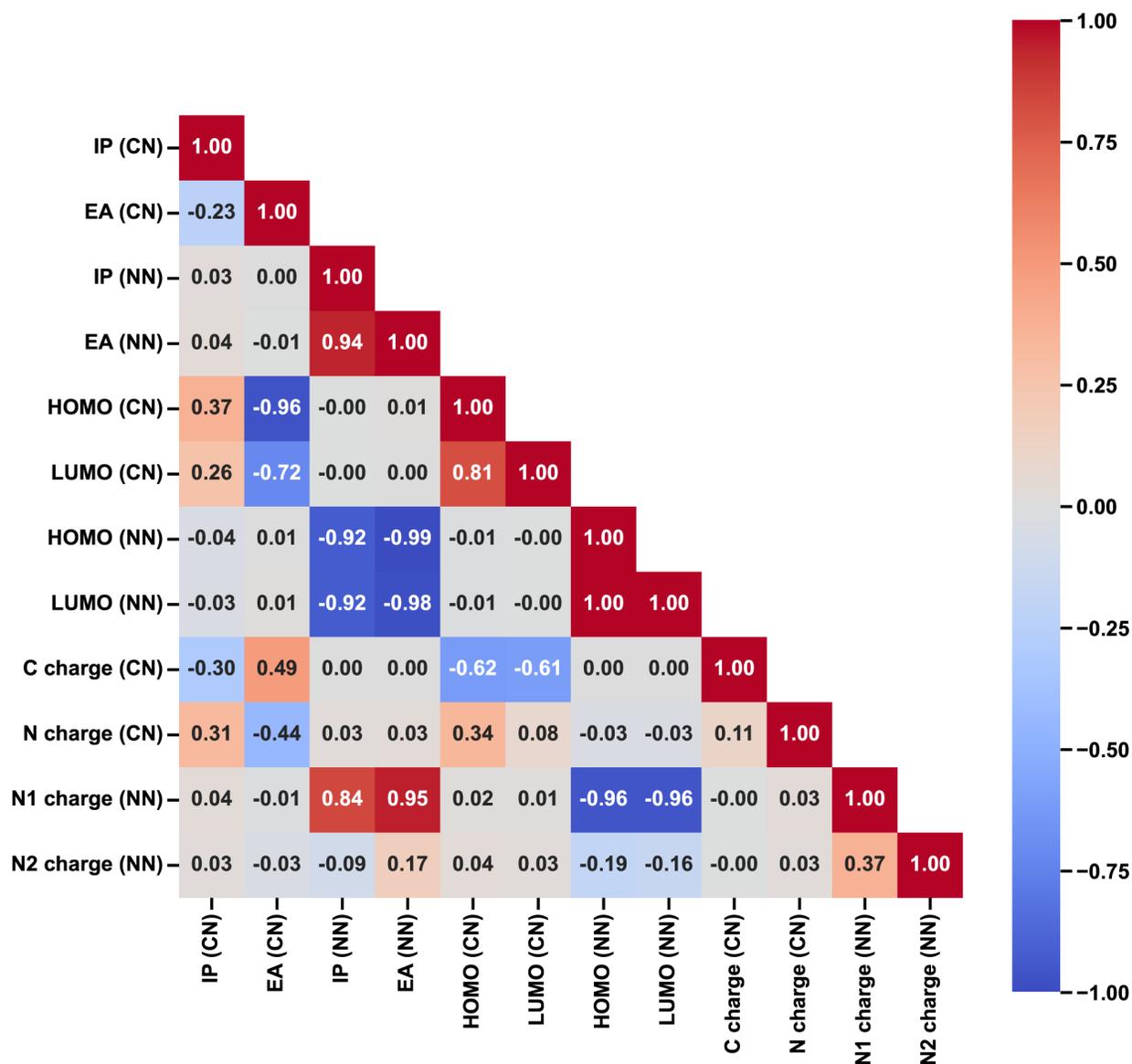

**Figure S2.** The signed Pearson correlation coefficients (-1 in blue to +1 in red with gray for 0 as indicated in the colorbar on the right) of the xTB features with each other across the original dataset in the experimental study of DiLuzio *et al.*[3], excluding the baseline *solvato* complexes that contain a DMSO ligand and complexes with a spectral integral below $10^5$ photon counts.

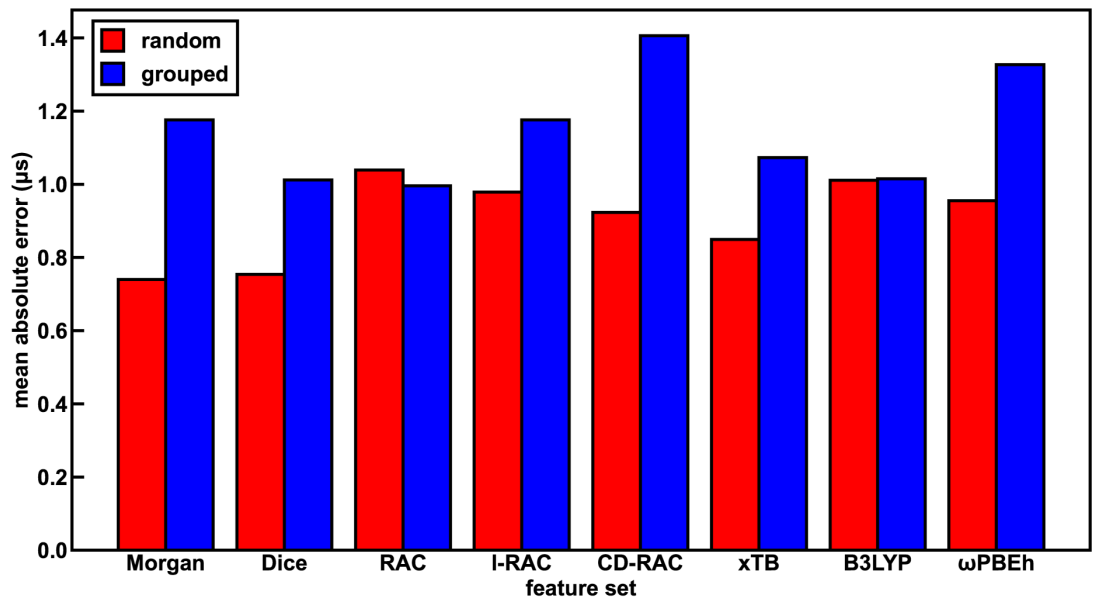

**Figure S3.** The test set performance of ANNs trained on different feature sets in predicting lifetime (in units of µs) for both random (red bars) and grouped splits (blue bars). The Morgan feature set leads to the best performance on the random split, and the RAC feature set leads to the best performance on the grouped split.

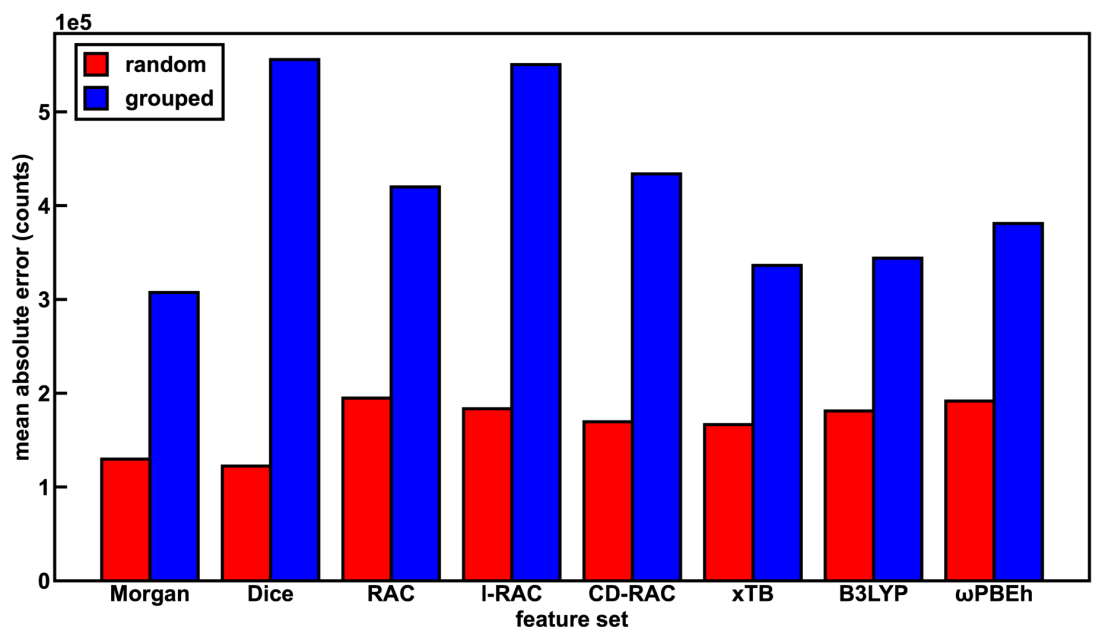

**Figure S4.** The test set performance of ANNs trained on different feature sets in predicting spectral integral (in units of photon counts) for both random (red bars) and grouped splits (blue bars). The Dice feature set leads to the best performance on the random split, and the Morgan feature set leads to the best performance on the grouped split.

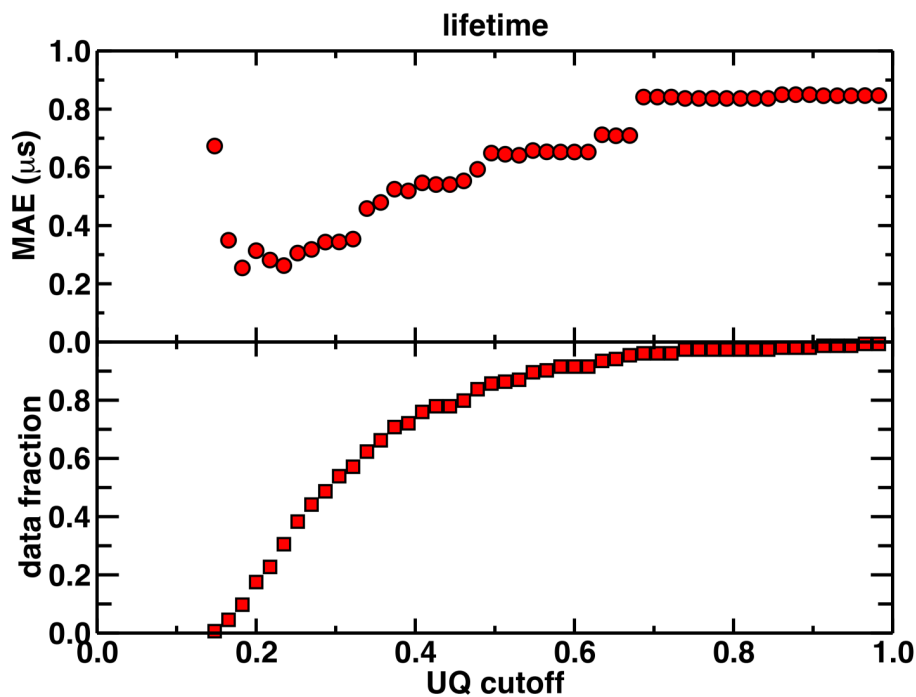

**Figure S5.** The uncertainty quantification (UQ) cutoff versus test set mean absolute error (in μs) and data fraction of the random split ANN model trained on the xTB feature set and predicting lifetime. The data fraction is the number of test set complexes under the corresponding UQ cutoff, and the MAE is calculated on this subset of complexes. The UQ metric used is the average latent space distance to the ten nearest neighbors in the training set following the protocol introduced in Ref. [8]. The UQ metric is normalized such that the largest UQ metric is scaled to 1.

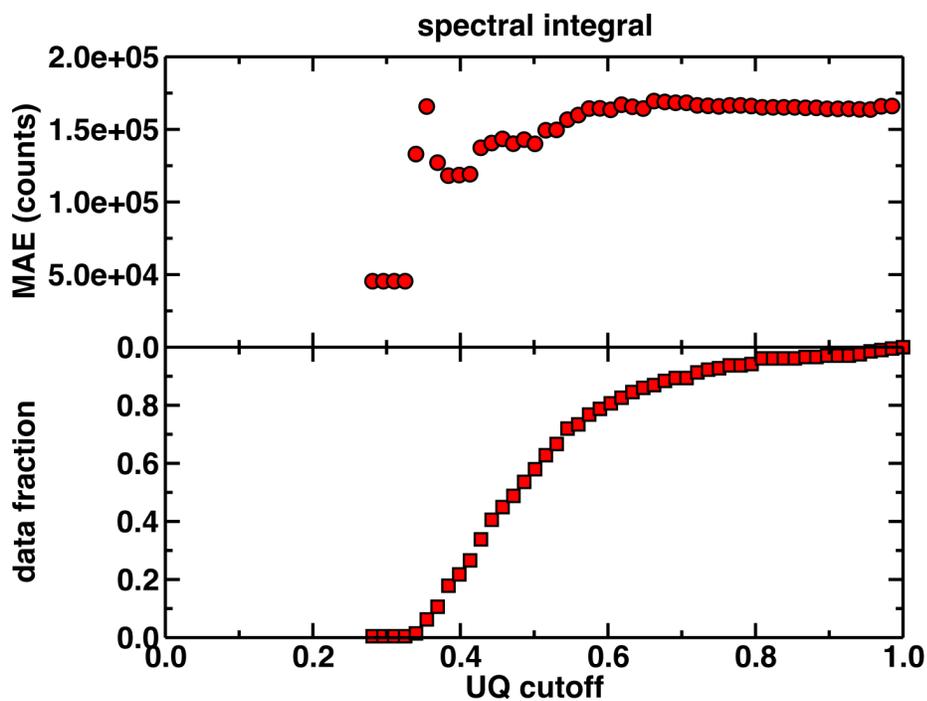

**Figure S6.** The uncertainty quantification (UQ) cutoff versus test set mean absolute error (in photon counts) of the random split ANN model trained on the xTB feature set and predicting spectral integral. The data fraction is the number of test set complexes under the corresponding UQ cutoff, and the MAE is calculated on this subset of complexes. The UQ metric used is the average latent space distance to the ten nearest neighbors in the training set following the protocol introduced in Ref. [8]. The UQ metric is normalized such that the largest UQ metric is scaled to 1.

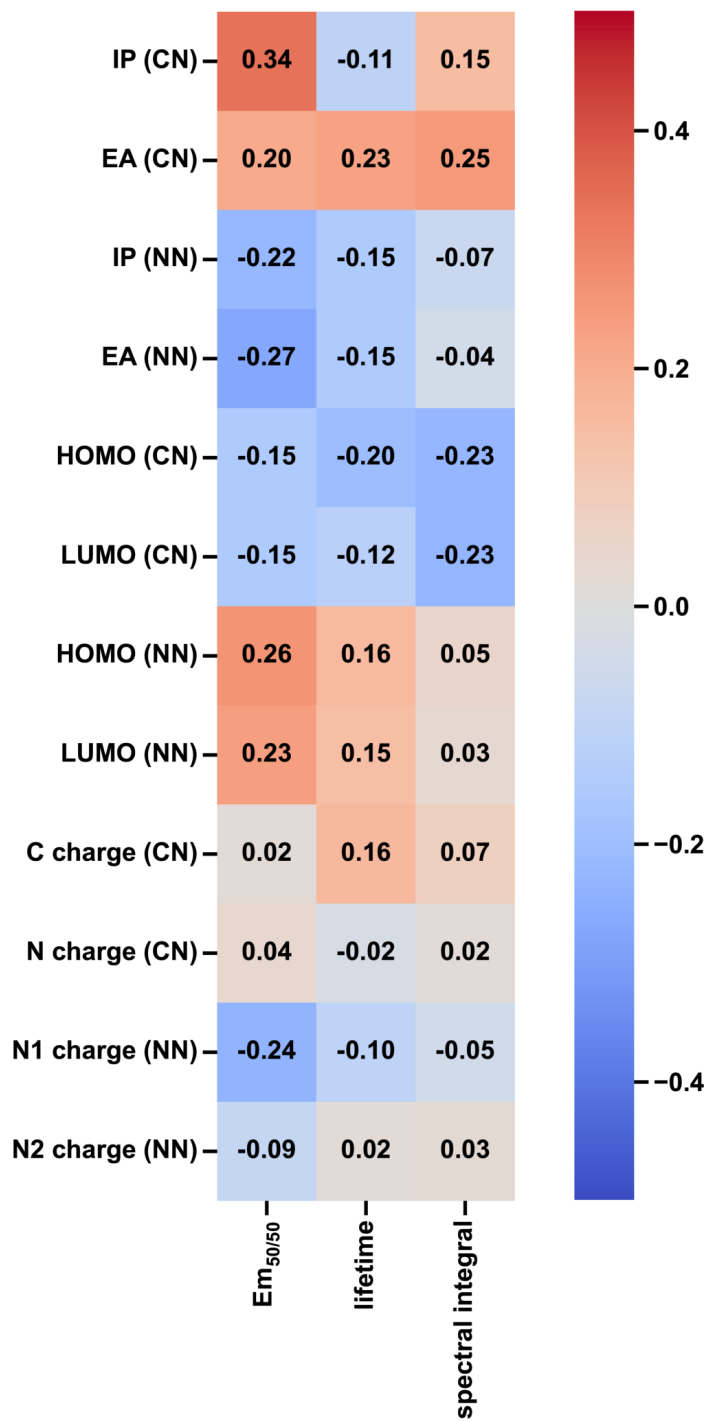

**Figure S7.** The signed Pearson correlation coefficients (-0.5 in blue to +0.5 in red with gray for 0 as indicated in the colorbar on the right) of the xTB features with the phosphor properties across the original dataset in the experimental study of DiLuzio *et al.*[3], excluding the baseline *solvato* complexes that contain a DMSO ligand and complexes with a spectral integral below $10^5$ photon counts.

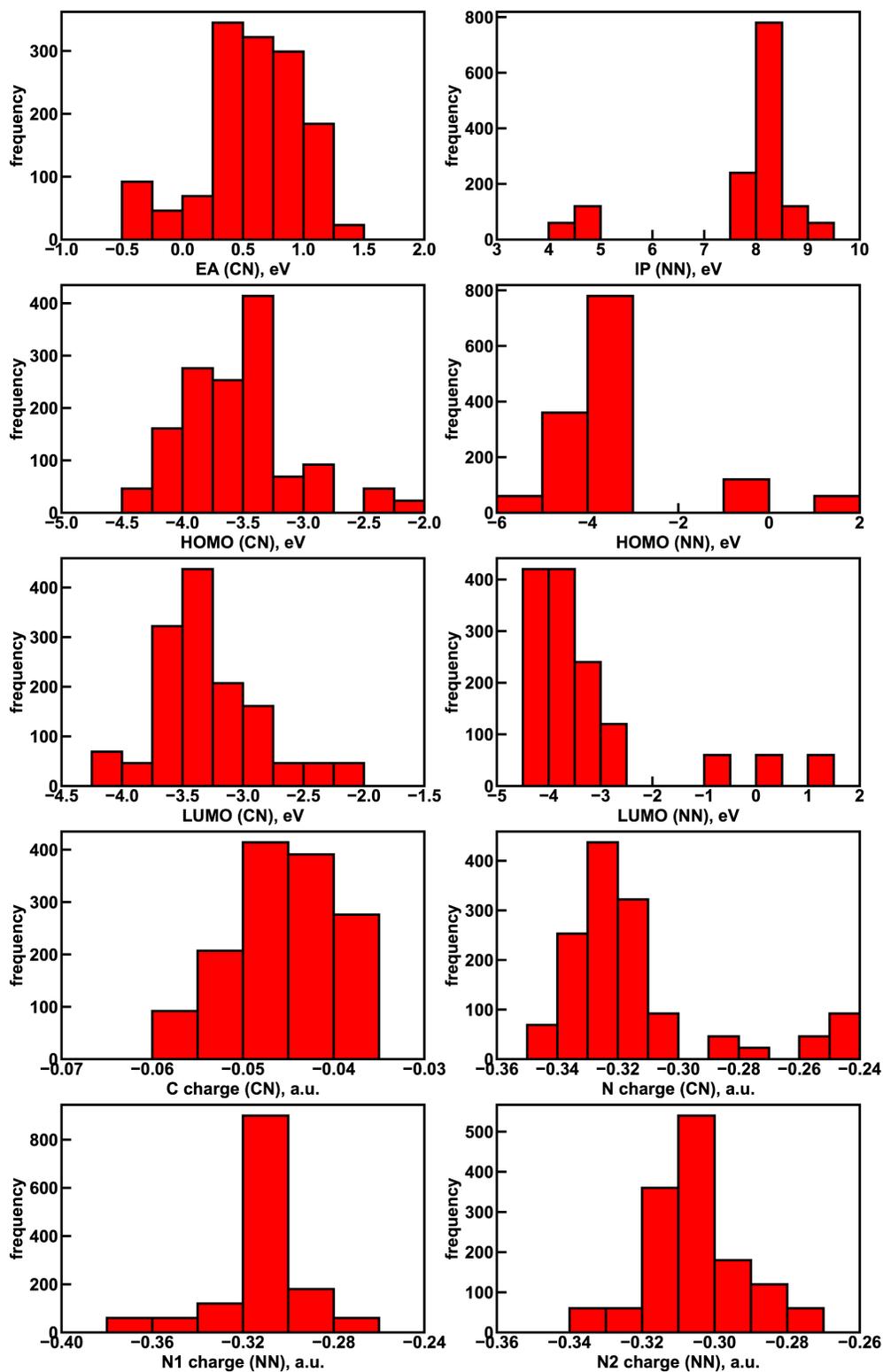

**Figure S8.** xTB feature distributions over the 1,380 complexes reported in the experimental study of DiLuzio *et al.*[3], excluding the baseline *solvato* complexes.

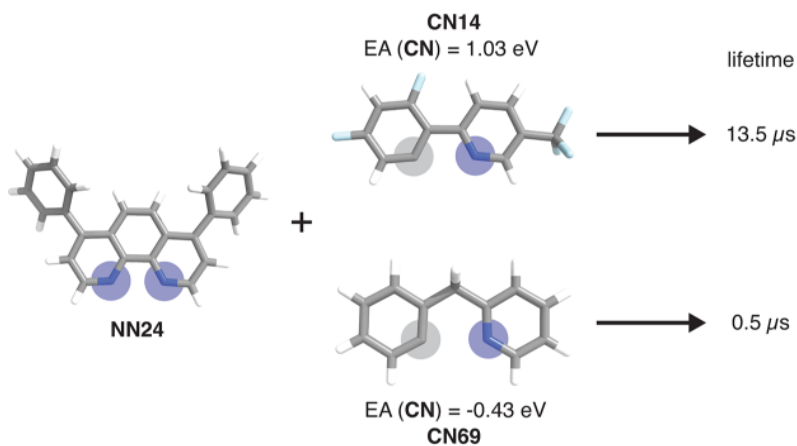

**Figure S9.** Example of a pair of complexes where the substitution of the **CN** ligand leads to a large lifetime property change. Here, complexes are represented by the combination of a **CN** and **NN** ligand. Coordinated nitrogen (carbon) atoms are indicated with blue (gray) circles. The relevant xTB features for the substituted ligands are shown. Atoms are colored as follows: white for hydrogen, gray for carbon, blue for nitrogen, and light blue for fluorine.

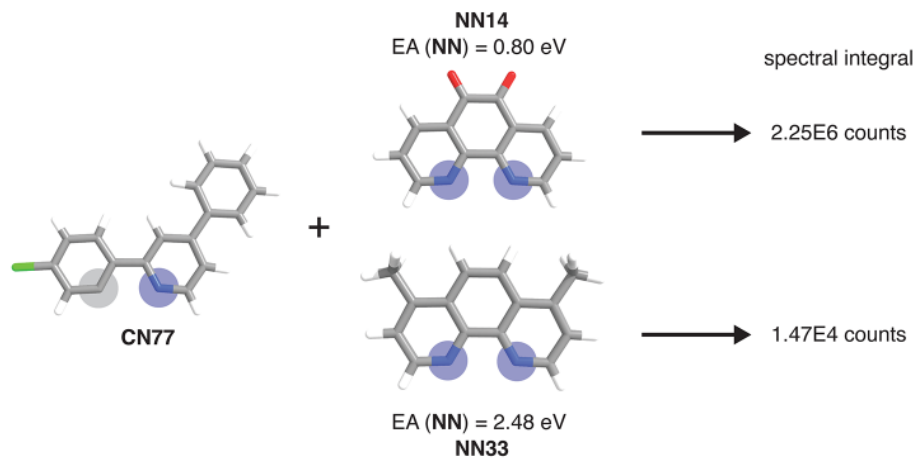

**Figure S10.** Example of a pair of complexes where the substitution of the **NN** ligand leads to a large spectral integral property change. Here, complexes are represented by the combination of a **CN** and **NN** ligand. Coordinated nitrogen (carbon) atoms are indicated with blue (gray) circles. The relevant xTB features for the substituted ligands are shown. Atoms are colored as follows: white for hydrogen, gray for carbon, blue for nitrogen, red for oxygen, and green for chlorine.

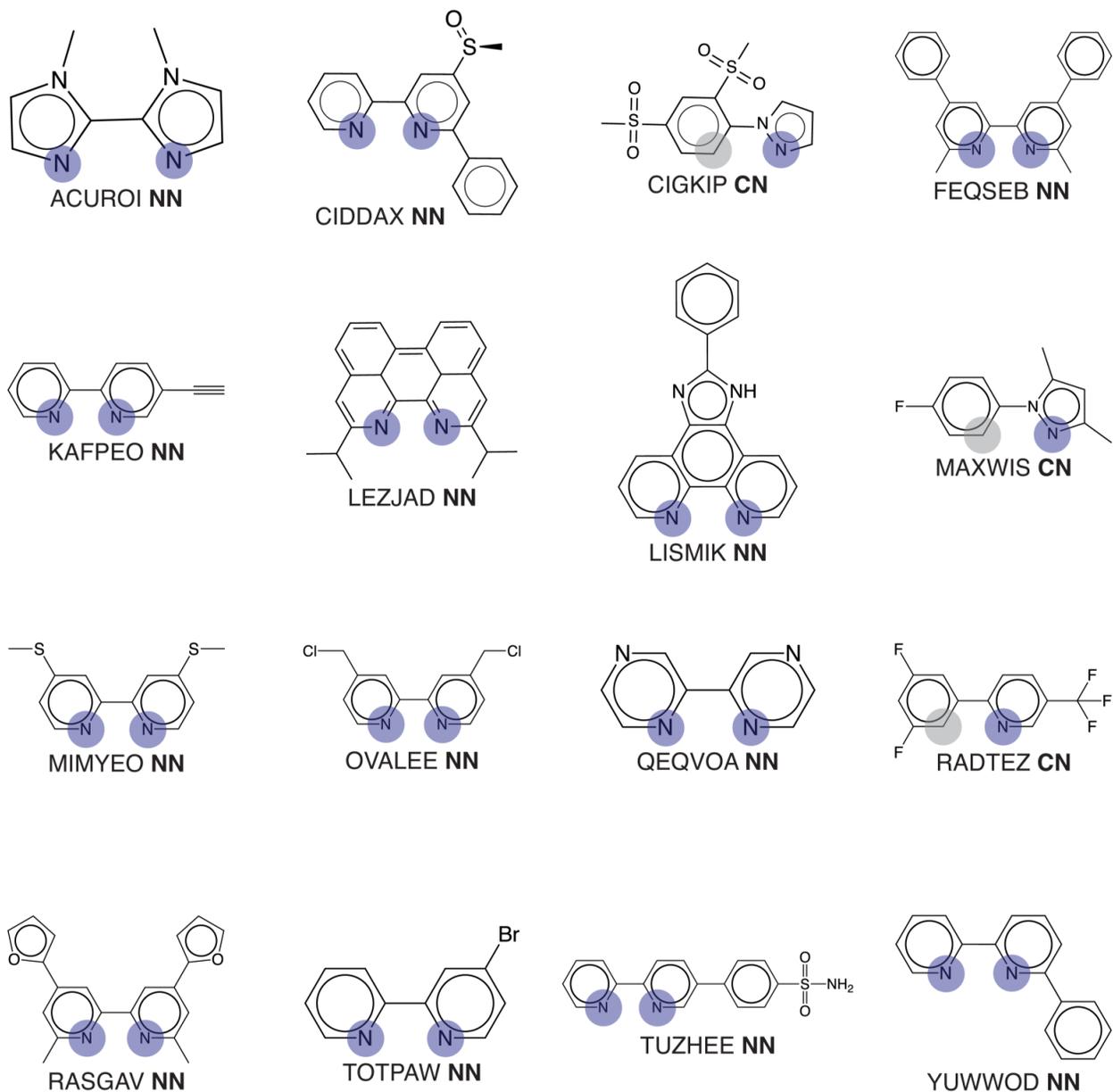

**Figure S11**. Sixteen CSD ligands that lead to extreme predicted phosphor properties. The six letter identifiers are CSD refcodes. Coordinated nitrogen (carbon) atoms are indicated with blue (gray) circles. ACUROI **NN** appears frequently in hypothetical complexes with high predicted spectral integral. CIDDAX **NN** is present in the hypothetical complex with the lowest predicted lifetime. CIGKIP **CN** is present in the hypothetical complex with the highest predicted spectral integral. FEQSEB **NN** appears frequently in hypothetical complexes with high predicted $Em_{50/50}$ and lifetime. KAFPEO **NN** appears frequently in hypothetical complexes with low predicted spectral integral. LEZJAD **NN** is present in two of the hypothetical complexes with the lowest predicted $Em_{50/50}$. LISMIK **NN** appears frequently in hypothetical complexes with low predicted $Em_{50/50}$. MAXWIS **CN** appears frequently in hypothetical complexes with low predicted lifetime and is present in the hypothetical complex with the third lowest predicted lifetime. MIMYEO **NN** appears frequently in hypothetical complexes with high predicted lifetime. OVALEE **NN** appears frequently in hypothetical complexes with low predicted lifetime. QEQVOA **NN** appears

frequently in hypothetical complexes with low predicted spectral integral. RADTEZ **CN** is present in the three hypothetical complexes with the highest predicted $Em_{50/50}$ and in the hypothetical complex with the second highest predicted spectral integral. RASGAV **NN** appears frequently in hypothetical complexes with high predicted $Em_{50/50}$ and lifetime and is present in the hypothetical complex with the third highest predicted spectral integral and the hypothetical complex with the third highest predicted lifetime. TOTPAW **NN** appears frequently in hypothetical complexes with low predicted $Em_{50/50}$ and lifetime and is present in the hypothetical complex with the highest predicted lifetime and the hypothetical complex with the second lowest predicted lifetime. TUZHEE **NN** appears frequently in hypothetical complexes with low predicted $Em_{50/50}$ and is present in the hypothetical complex with the second lowest predicted $Em_{50/50}$. YUWWOD **NN** appears frequently in hypothetical complexes with high predicted spectral integral and $Em_{50/50}$ and is present in the hypothetical complex with the third highest predicted $Em_{50/50}$ and the hypothetical complex with the second highest predicted lifetime.

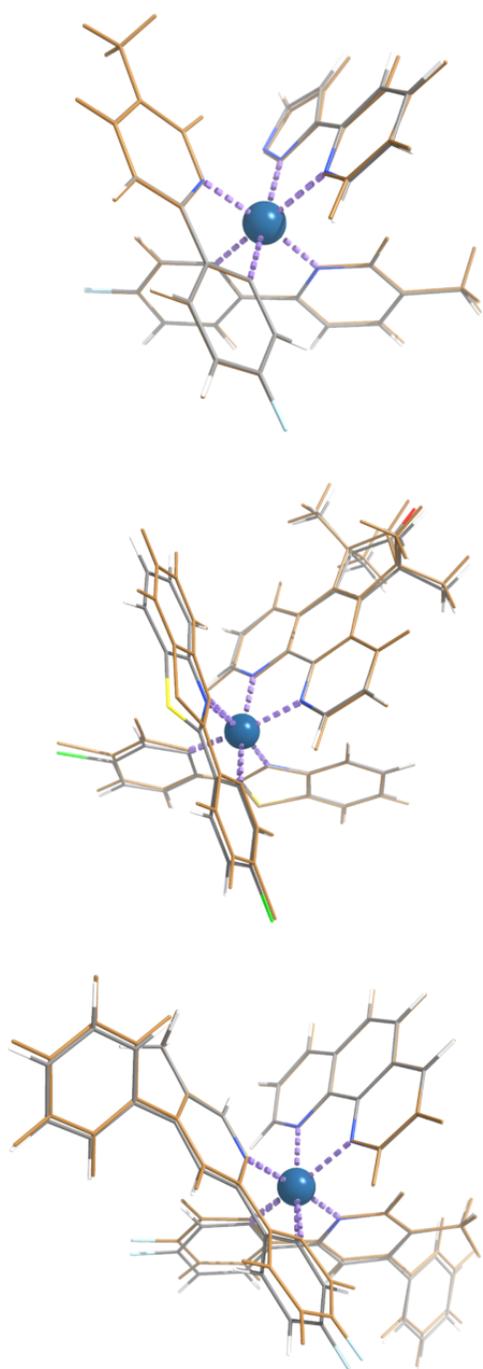

**Figure S12.** Comparison of singlet and triplet B3LYP geometries of select iridium complexes, [Ir(**CN3**)$_2$(**NN40**)]$^0$ (top), [Ir(**CN38**)$_2$(**NN27**)]$^+$ (middle), and [Ir(**CN75**)$_2$(**NN16**)]$^+$ (bottom). Triplet geometries are shown in brown, while singlet geometries are colored normally with white for hydrogen, gray for carbon, blue for nitrogen, red for oxygen, light blue for fluorine, yellow for sulfur, green for chlorine, and dark blue for iridium. The RMSD values between these singlet-triplet structure pairs are 0.065 Angstroms, 0.188 Angstroms, and 0.164 Angstroms, respectively.

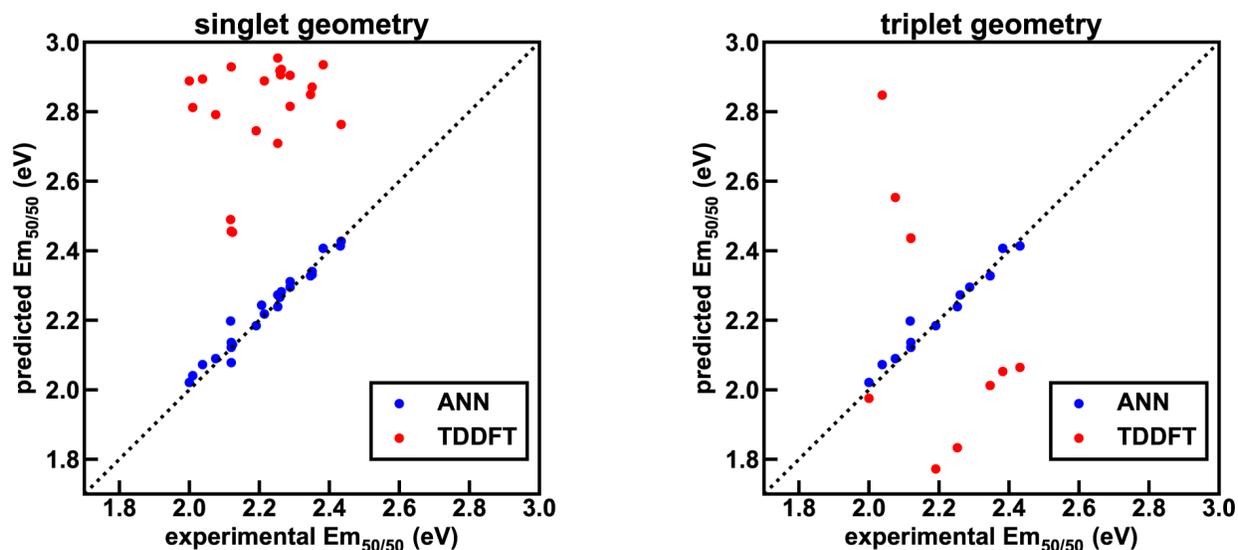

**Figure S13.** TDDFT emission energy predictions with the CAM-B3LYP functional, on iridium phosphor geometries optimized with the CAM-B3LYP functional in the singlet (left) and triplet (right) state.

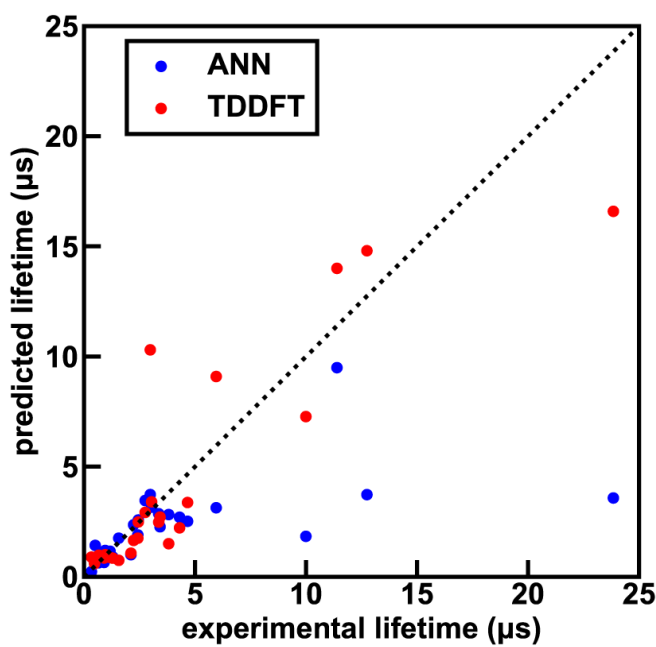

**Figure S14.** Comparison of random split ANN and TDDFT lifetime prediction to experiment (in μs) across 26 test set iridium complexes in the experimental dataset. These complexes were chosen to span the range of emission energies and lifetimes of the full set. TDDFT was carried out on optimized singlet geometries using the B3LYP functional. Lifetime was calculated using excitation energy and transition dipole moment output from the TDDFT calculation (see Methods). The dotted line is included as a reference and corresponds to perfect agreement between prediction and experiment.

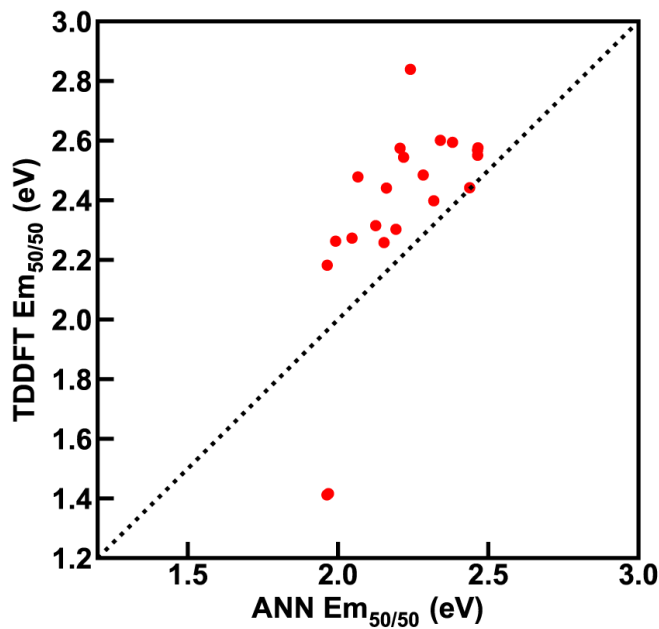

**Figure S15.** Comparison of random split ANN and TDDFT $Em_{50/50}$ predictions (in eV) across 21 hypothetical iridium complexes. TDDFT was carried out on optimized singlet geometries using the B3LYP functional. The energy of the three lowest triplet sublevels was averaged for the TDDFT energy, which is used to approximate $Em_{50/50}$.

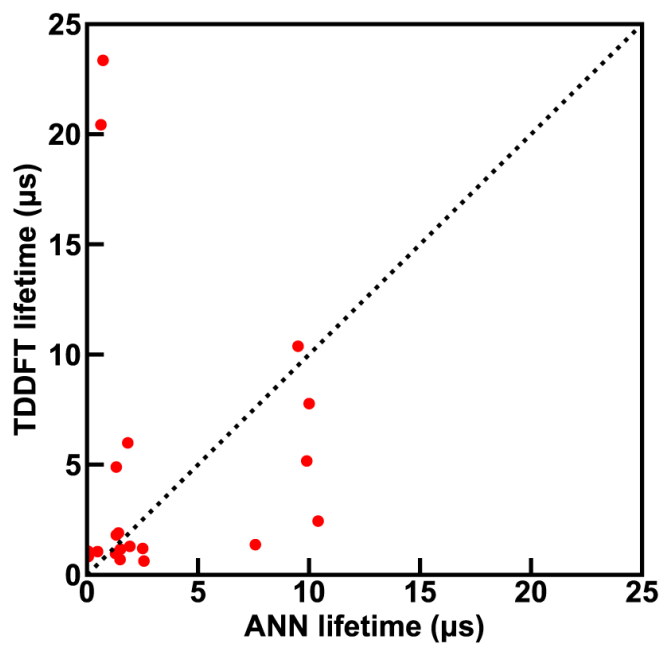

**Figure S16.** Comparison of random split ANN and TDDFT lifetime predictions (in μs) across 21 hypothetical iridium complexes. TDDFT was carried out on optimized singlet geometries using the B3LYP functional. Lifetime was calculated using excitation energy and transition dipole moment output from the TDDFT calculation (see Methods).

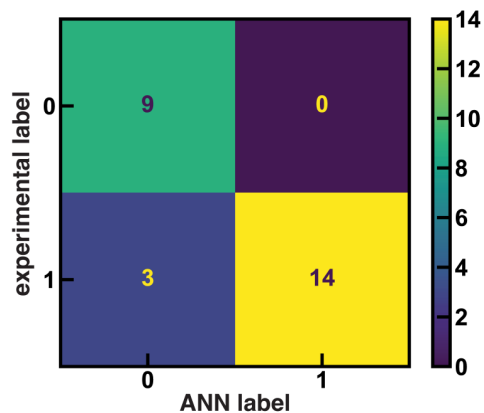
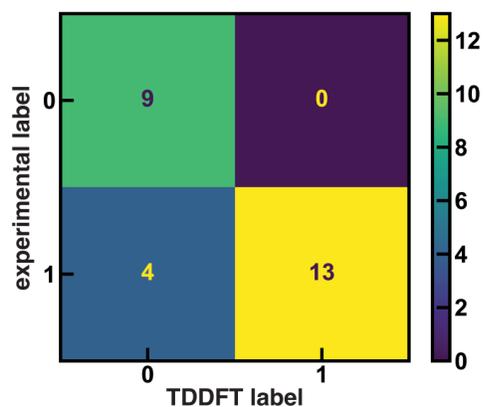
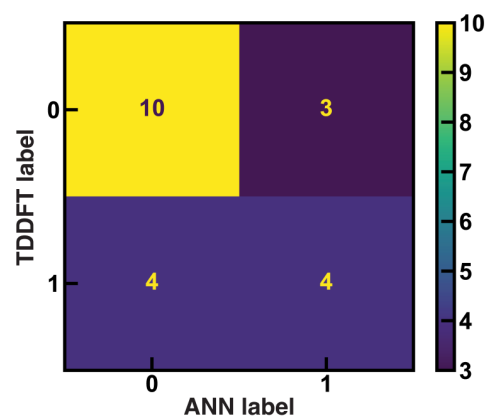

**Figure S17.** Confusion matrices indicating the agreement between different lifetime labels. 0 corresponds to a lifetime $\tau \leq 2$ μs and 1 corresponds to $\tau > 2$ μs. Over the 26 representative test set experimental complexes, we show experimentally measured lifetimes versus random split ANN predictions (top) and TDDFT predictions (middle). Over the 21 hypothetical complexes, we show TDDFT predictions versus random split ANN predictions (bottom). The TDDFT results were generated with singlet geometries and the B3LYP functional.

**Table S1.** The labels and structures of the cyclometalating (**CN**) ligands in the HLS. XYZ files of these ligands are provided in the Electronic Supplementary Information .zip file.

| | | | |
|---|---|---|---|
| **CN1** | 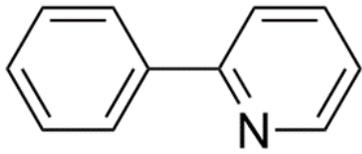 | **CN63** | 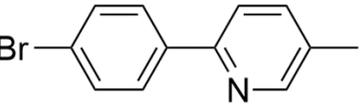 |
| **CN2** | 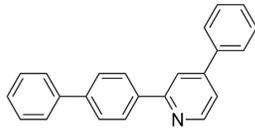 | **CN64** | 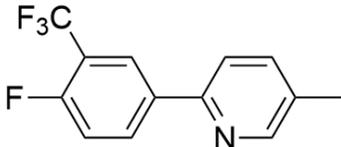 |
| **CN3** | 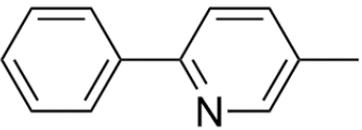 | **CN65** | 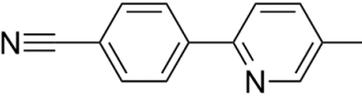 |
| **CN4** | 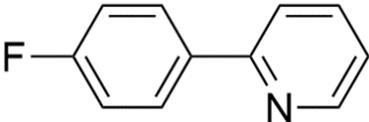 | **CN66** | 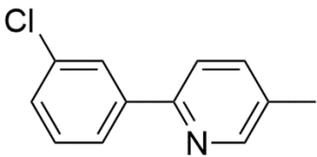 |
| **CN5** | 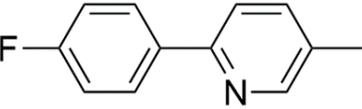 | **CN67** | 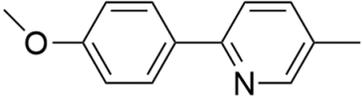 |
| **CN7** | 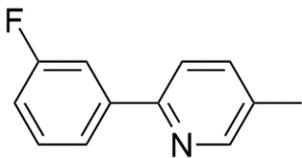 | **CN68** | 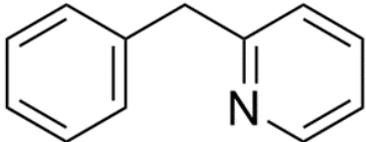 |
| **CN9** | 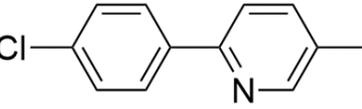 | **CN69** | 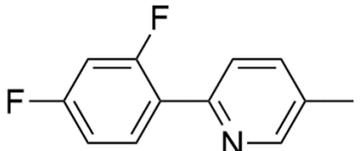 |
| **CN11** | 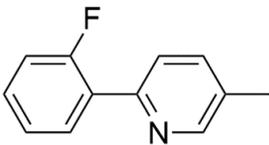 | **CN70** | 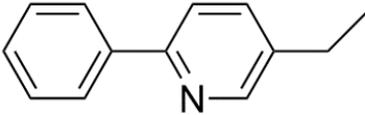 |
| **CN12** | 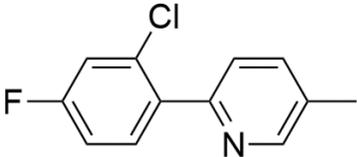 | **CN71** | 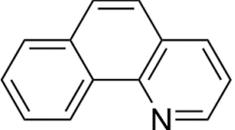 |
| **CN13** | 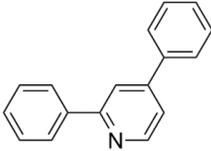 | **CN72** | 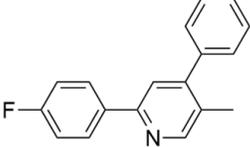 |
| **CN14** | 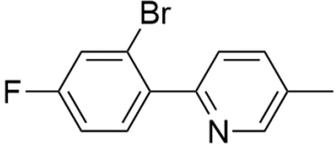 | **CN73** | 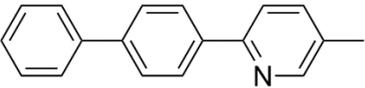 |
| **CN21** | 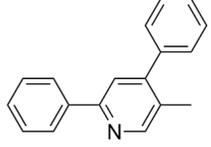 | **CN74** | 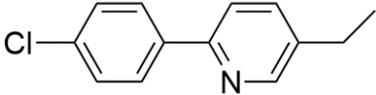 |
| **CN28** | 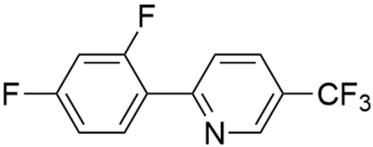 | **CN75** | 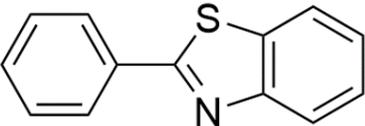 |
| **CN29** | 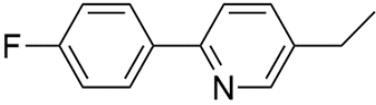 | **CN76** | 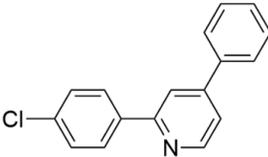 |
| **CN30** | 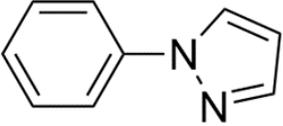 | **CN77** | 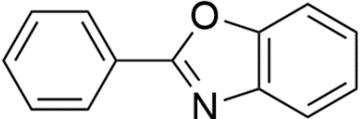 |
| **CN31** | 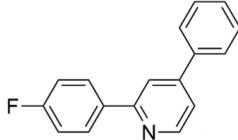 | **CN78** | 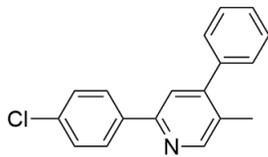 |
| **CN33** | 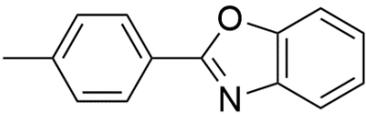 | **CN79** | 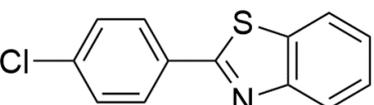 |
| **CN34** | 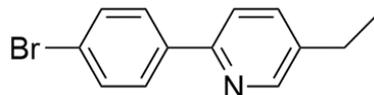 | **CN80** | 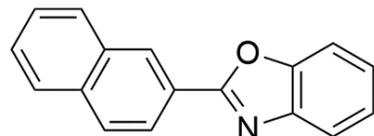 |
| **CN35** | 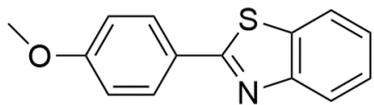 | **CN81** | 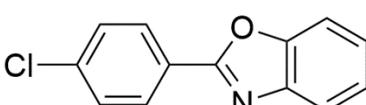 |
| **CN37** | 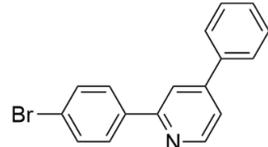 | **CN94** | 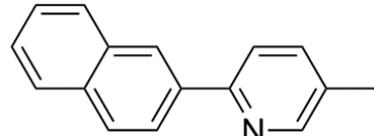 |
| **CN38** | 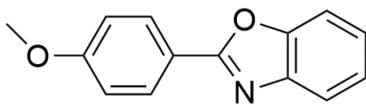 | **CN95** | 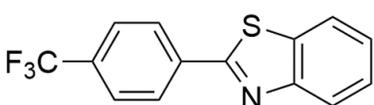 |
| **CN39** | 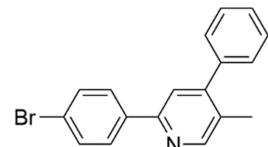 | **CN101** | 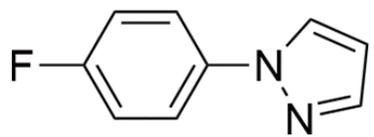 |
| **CN40** | 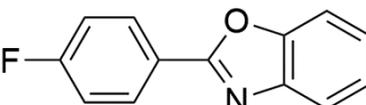 | **CN102** | 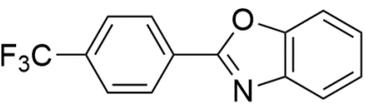 |
| **CN41** | 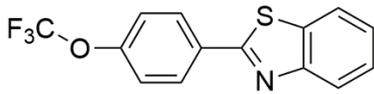 | **CN103** | 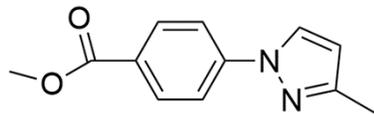 |
| **CN42** | 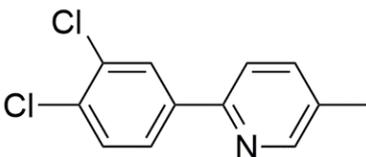 | **CN104** | 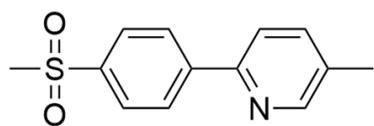 |
| **CN44** | 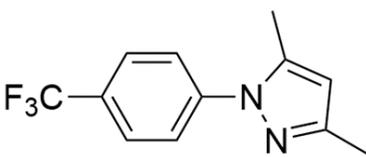 | **CN105** | 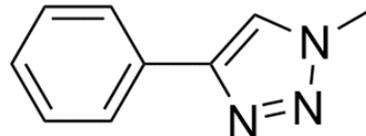 |
| **CN46** | 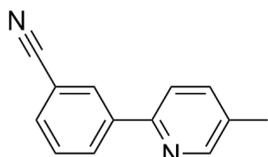 | **CN106** | 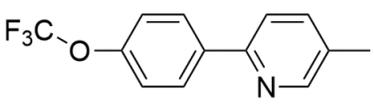 |
| **CN48** | 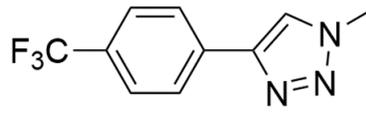 | **CN107** | 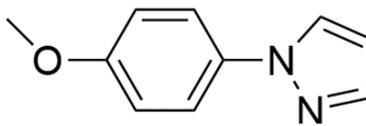 |
| **CN49** | 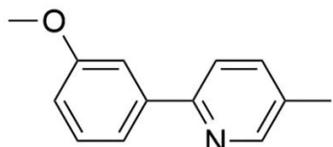 | **CN108** | 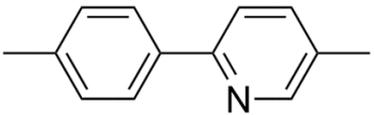 |
| **CN54** | 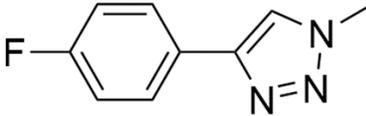 | **CN109** | 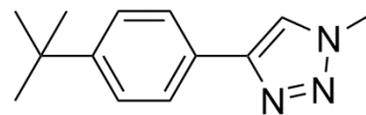 |

**Table S2.** The labels and structures of the ancillary (**NN**) ligands in the HLS. XYZ files of these ligands are provided in the Electronic Supplementary Information .zip file. In the .zip file, **NN40**, **NN41**, and **NN42** are in deprotonated form.

| | | | |
|---|---|---|---|
| **NN1** | 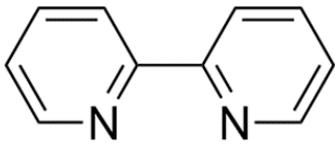 | **NN24** | 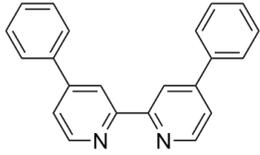 |
| **NN2** | 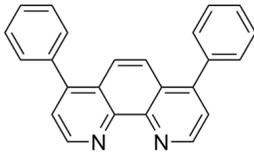 | **NN26** | 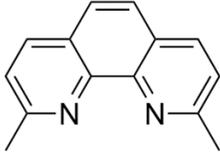 |
| **NN3** | 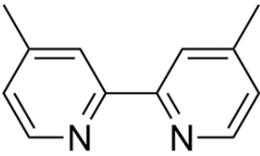 | **NN27** | 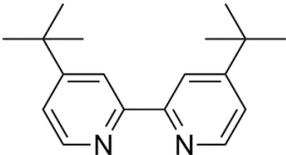 |
| **NN4** | 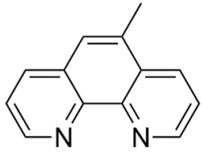 | **NN33** | 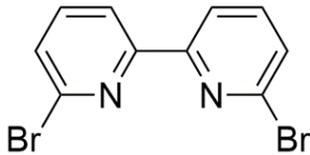 |
| **NN5** | 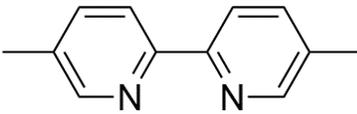 | **NN34** | 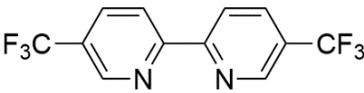 |
| **NN6** | 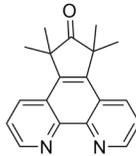 | **NN37** | 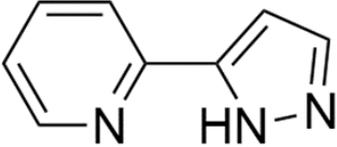 |
| **NN7** | 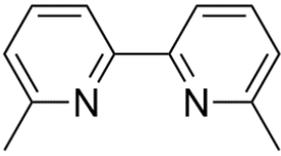 | **NN40** | 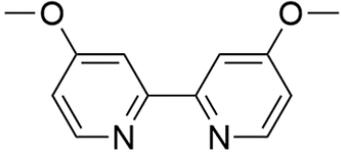 |
| **NN8** | 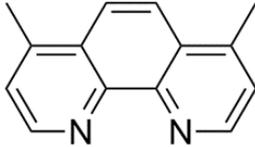 | **NN41** | 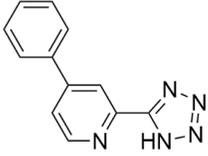 |
| **NN14** | 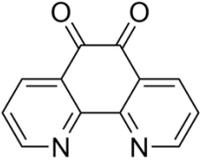 | **NN42** | 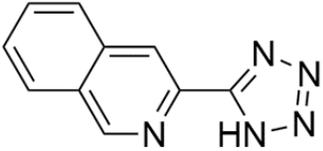 |
| **NN16** | 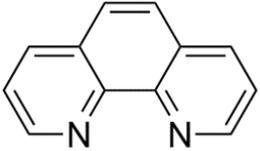 | **NN43** | 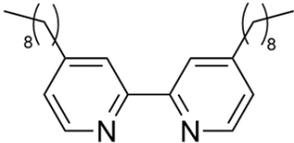 |
| **NN20** | 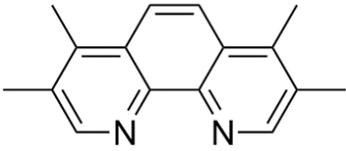 | **NN47** | 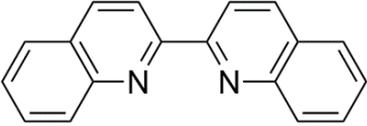 |
| **NN21** | 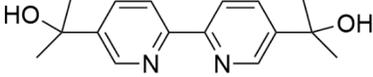 | | |

**Table S3.** The eighty-three random split features in the Dice feature set. Features and notation are described in detail in Text S2. For the grouped split, there are five fewer features in the Dice feature set due to the absence of five HLS ligands from the training data.

| feature type | feature count | feature list |
|---|---|---|
| Dice coefficient between the Morgan fingerprint of the **CN** ligand of the current complex and the Morgan fingerprint of an HLS **CN** ligand | 60 | similarity to **CN1** similarity to **CN2** similarity to **CN3** … similarity to **CN109** |
| Dice coefficient between the Morgan fingerprint of the **NN** ligand of the current complex and the Morgan fingerprint of an HLS **NN** ligand | 23 | similarity to **NN1** similarity to **NN2** similarity to **NN3** … similarity to **NN47** |

**Table S4.** Comparison of random split ANN performance, as measured by the mean absolute errors of model test set predictions, using the Dice feature set versus using an analogous Tanimoto feature set. We used RDKit 2021.9.2[9] to generated Morgan fingerprints of ligands and Tanimoto similarity coefficients in the same way Dice similarity coefficients were generated (see main text *Feature Sets* discussion).

|  | target property | | |
|---|---|---|---|
|  | $Em_{50/50}$ (eV) | lifetime (μs) | spectral integral (photon counts) |
| Dice | 0.0161 | 0.7540 | 1.22E5 |
| Tanimoto | 0.0169 | 0.8381 | 1.31E5 |

**Table S5**. The 196 features in the standard RAC feature set. Invariant features are removed from the original twenty-eight features of that category (seven atomic properties and four depths, i.e. 0, 1, 2, and 3). Features and notation are described in detail in Text S2.

| start | extent | operation | features removed | feature count |
|---|---|---|---|---|
| full-scope | all | product | 0 | 28 |
|  | axial ligand | product | 0 | 28 |
|  | equatorial ligand | product | 0 | 28 |
| metal-centered | all | product | (14) mc-chi-0-all, mc-Z-0-all, mc-Z-1-all, mc-I-0-all, mc-I-1-all, mc-I-2-all, mc-T-0-all, mc-T-1-all, mc-S-0-all, mc-S-1-all, mc-Gval-0-all, mc-Gval-1-all, mc-NumB-0-all, mc-NumB-1-all | 14 |
|  |  | difference | (16) D_mc-chi-0-all, D_mc-chi-1-all, D_mc-Z-0-all, D_mc-Z-1-all, D_mc-I-0-all, D_mc-I-1-all, D_mc-I-2-all, D_mc-I-3-all, D_mc-T-0-all, D_mc-T-1-all, D_mc-S-0-all, D_mc-S-1-all, D_mc-Gval-0-all, D_mc-Gval-1-all, D_mc-NumB-0-all, D_mc-NumB-1-all | 12 |
| ligand-centered | axial ligand | product | (3) lc-I-0-ax, lc-I-1-ax, lc-T-0-ax | 25 |
|  |  | difference | (10) D_lc-chi-0-ax, D_lc-Z-0-ax, D_lc-I-0-ax, D_lc-I-1-ax, D_lc-I-2-ax, D_lc-I-3-ax, D_lc-T-0-ax, D_lc-S-0-ax, D_lc-Gval-0-ax, D_lc-NumB-0-ax | 18 |
|  | equatorial ligand | product | (3) lc-I-0-eq, lc-I-1-eq, lc-T-0-eq | 25 |
|  |  | difference | (10) D_lc-chi-0-eq, D_lc-Z-0-eq, D_lc-I-0-eq, D_lc-I-1-eq, D_lc-I-2-eq, D_lc-I-3-eq, D_lc-T-0-eq, D_lc-S-0-eq, D_lc-Gval-0-eq, D_lc-NumB-0-eq | 18 |

**Table S6.** The seventy features in the ligand-only RAC feature set. Each combination of ligand and atomic property yields five features due to depths ranging from 0 to 4. Features and notation are described in detail in Text S2.

| ligand | atomic property | feature list |
|---|---|---|
| **CN** | topology | T-0_CN, T-1_CN, T-2_CN, T-3_CN, T-4_CN |
|  | identity | I-0_CN, I-1_CN, I-2_CN, I-3_CN, I-4_CN |
|  | electronegativity | chi-0_CN, chi-1_CN, chi-2_CN, chi-3_CN, chi-4_CN |
|  | covalent radius | S-0_CN, S-1_CN, S-2_CN, S-3_CN, S-4_CN |
|  | nuclear charge | Z-0_CN, Z-1_CN, Z-2_CN, Z-3_CN, Z-4_CN |
|  | group number | Gval-0_CN, Gval-1_CN, Gval-2_CN, Gval-3_CN, Gval-4_CN |
|  | number of bonds | NumB-0_CN, NumB-1_CN, NumB-2_CN, NumB-3_CN, NumB-4_CN |
| **NN** | topology | T-0_NN, T-1_NN, T-2_NN, T-3_NN, T-4_NN |
|  | identity | I-0_NN, I-1_NN, I-2_NN, I-3_NN, I-4_NN |
|  | electronegativity | chi-0_NN, chi-1_NN, chi-2_NN, chi-3_NN, chi-4_NN |
|  | covalent radius | S-0_NN, S-1_NN, S-2_NN, S-3_NN, S-4_NN |
|  | nuclear charge | Z-0_NN, Z-1_NN, Z-2_NN, Z-3_NN, Z-4_NN |
|  | group number | Gval-0_NN, Gval-1_NN, Gval-2_NN, Gval-3_NN, Gval-4_NN |
|  | number of bonds | NumB-0_NN, NumB-1_NN, NumB-2_NN, NumB-3_NN, NumB-4_NN |

**Table S7**. The 222 features in the CD-RAC feature set. Invariant features are removed from the original twenty-eight features of that category (seven atomic properties and four depths, i.e. 0, 1, 2, and 3). The notation of features and property type is the same as in RACs but incorporates a distance-dependent term as outlined in ref. [10]. Features and notation are described in detail in Text S2.

| start | extent | operation | features removed | feature count |
|---|---|---|---|---|
| full-scope | all | product | 0 | 28 |
| | axial ligand | product | 0 | 28 |
| | equatorial ligand | product | 0 | 28 |
| metal-centered | all | product | 0 | 28 |
| | | difference | (10) D_mc-chi-0-all, D_mc-Z-0-all, D_mc-I-0-all, D_mc-I-1-all, D_mc-I-2-all, D_mc-I-3-all, D_mc-T-0-all, D_mc-S-0-all, D_mc-Gval-0-all, D_mc-NumB-0-all | 18 |
| ligand-centered | axial ligand | product | 0 | 28 |
| | | difference | (10) D_lc-chi-0-ax, D_lc-Z-0-ax, D_lc-I-0-ax, D_lc-I-1-ax, D_lc-I-2-ax, D_lc-I-3-ax, D_lc-T-0-ax, D_lc-S-0-ax, D_lc-Gval-0-ax, D_lc-NumB-0-ax | 18 |
| | equatorial ligand | product | 0 | 28 |
| | | difference | (10) D_lc-chi-0-eq, D_lc-Z-0-eq, D_lc-I-0-eq, D_lc-I-1-eq, D_lc-I-2-eq, D_lc-I-3-eq, D_lc-T-0-eq, D_lc-S-0-eq, D_lc-Gval-0-eq, D_lc-NumB-0-eq | 18 |

**Table S8.** The twelve features present in each of the xTB, B3LYP DFT, and ωPBEh DFT feature sets (i.e., evaluated at each level of theory).

| feature type | ligand | feature count | feature list |
|---|---|---|---|
| energy levels | **CN** | 2 | HOMO of **CN** ligand<br>LUMO of **CN** ligand |
| | **NN** | 2 | HOMO of **NN** ligand<br>LUMO of **NN** ligand |
| energy descriptor | **CN** | 2 | IP of **CN** ligand<br>EA of **CN** ligand |
| | **NN** | 2 | IP of **NN** ligand<br>EA of **NN** ligand |
| Mulliken charges | **CN** | 2 | charge of coordinating carbon<br>charge of coordinating nitrogen |
| | **NN** | 2 | charge of first coordinating nitrogen<br>charge of second coordinating nitrogen |

**Table S9.** Comparison of random split ANN performance, as measured by the mean absolute errors of model test set predictions, using natural charges instead of Mulliken charges for the connecting atom charge features of the B3LYP DFT and ωPBEh DFT feature sets. Although Mulliken charges lead to worse performance on the random split, they lead to better performance on the grouped split. Regardless of which charge scheme is used for the DFT feature sets, the xTB feature set outperforms the B3LYP DFT and ωPBEh DFT feature sets.

|  |  | target property | | |
|---|---|---|---|---|
|  |  | $Em_{50/50}$ (eV) | lifetime (μs) | spectral integral (photon counts) |
| B3LYP DFT | Mulliken charges | 0.0290 | 1.0105 | 1.81E5 |
|  | natural charges | 0.0251 | 1.0134 | 1.75E5 |
| ωPBEh DFT | Mulliken charges | 0.0285 | 0.9550 | 1.92E5 |
|  | natural charges | 0.0233 | 0.9659 | 1.69E5 |

**Table S10.** The ranking of random split ANNs using different feature sets in predicting for each of the target properties. The ranking is on the basis of test set MAE, and a rank of 1 indicates the corresponding feature set led to the ANN with the lowest MAE for the specified target property.

|  | $Em_{50/50}$ rank | lifetime rank | spectral integral rank | mean rank |
|---|---|---|---|---|
| Dice | 1 | 2 | 1 | 1.33 |
| Morgan | 2 | 1 | 2 | 1.67 |
| xTB | 3 | 3 | 3 | 3.00 |
| CD-RAC | 5 | 4 | 4 | 4.33 |
| ligand-only RAC | 4 | 6 | 6 | 5.33 |
| ωPBEh DFT | 7 | 5 | 7 | 6.33 |
| B3LYP DFT | 8 | 7 | 5 | 6.67 |
| RAC | 6 | 8 | 8 | 7.33 |

**Table S11.** The MAE, scaled MAE, and $R^2$ test set performance of the eight random split ANNs predicting for $Em_{50/50}$. Scaled MAE is defined as MAE divided by the difference between the maximum and minimum value of the target property in the training data.

|  | MAE (eV) | scaled MAE | $R^2$ |
|---|---|---|---|
| Morgan | 0.0163 | 0.031 | 0.91 |
| Dice | 0.0161 | 0.031 | 0.92 |
| RAC | 0.0274 | 0.053 | 0.83 |
| ligand-only RAC | 0.0220 | 0.042 | 0.89 |
| CD-RAC | 0.0233 | 0.045 | 0.87 |
| xTB | 0.0210 | 0.040 | 0.91 |
| B3LYP DFT | 0.0290 | 0.056 | 0.83 |
| ωPBEh DFT | 0.0285 | 0.055 | 0.85 |

**Table S12.** The MAE, scaled MAE, and $R^2$ test set performance of the eight random split ANNs predicting for lifetime.

|  | MAE (μs) | scaled MAE | $R^2$ |
|---|---|---|---|
| Morgan | 0.7397 | 0.045 | 0.36 |
| Dice | 0.7540 | 0.046 | 0.38 |
| RAC | 1.0389 | 0.064 | 0.26 |
| ligand-only RAC | 0.9789 | 0.060 | 0.20 |
| CD-RAC | 0.9233 | 0.057 | 0.32 |
| xTB | 0.8495 | 0.052 | 0.38 |
| B3LYP DFT | 1.0105 | 0.062 | 0.29 |
| ωPBEh DFT | 0.9550 | 0.059 | 0.33 |

**Table S13.** The MAE, scaled MAE, and $R^2$ test set performance of the eight random split ANNs predicting for spectral integral.

|  | MAE (photon counts) | scaled MAE | $R^2$ |
|---|---|---|---|
| Morgan | 1.30E+05 | 0.054 | 0.84 |
| Dice | 1.22E+05 | 0.050 | 0.87 |
| RAC | 1.95E+05 | 0.080 | 0.73 |
| ligand-only RAC | 1.83E+05 | 0.075 | 0.73 |
| CD-RAC | 1.70E+05 | 0.070 | 0.79 |
| xTB | 1.67E+05 | 0.068 | 0.79 |
| B3LYP DFT | 1.81E+05 | 0.074 | 0.77 |
| ωPBEh DFT | 1.92E+05 | 0.079 | 0.72 |

Table S14. The Pearson correlation coefficient for the analogous electronic structure features in the xTB feature set and the B3LYP DFT and ωPBEh DFT feature sets. The correlations are evaluated across the original dataset in the experimental study of DiLuzio et al.[3], excluding the baseline *solvato* complexes that contain a DMSO ligand.

|  | xTB vs B3LYP | xTB vs ωPBEh |
|---|---|---|
| IP (**CN**) | 0.93 | 0.50 |
| EA (**CN**) | 0.96 | 0.93 |
| IP (**NN**) | 1.00 | 0.90 |
| EA (**NN**) | 0.99 | 0.98 |
| HOMO (**CN**) | 0.13 | 0.05 |
| LUMO (**CN**) | 0.56 | 0.55 |
| HOMO (**NN**) | 0.94 | 0.86 |
| LUMO (**NN**) | 0.91 | 0.93 |
| C charge (**CN**) | -0.19 | -0.13 |
| N charge (**CN**) | 0.86 | 0.86 |
| N1 charge (**NN**) | -0.05 | 0.05 |
| N2 charge (**NN**) | -0.15 | -0.06 |

**Table S15.** The MAE, scaled MAE, and $R^2$ test set performance of the eight grouped split ANNs predicting for $Em_{50/50}$.

|  | MAE (eV) | scaled MAE | $R^2$ |
|---|---|---|---|
| Morgan | 0.0522 | 0.100 | 0.41 |
| Dice | 0.0722 | 0.138 | -0.27 |
| RAC | 0.0547 | 0.105 | 0.46 |
| ligand-only RAC | 0.0646 | 0.124 | 0.17 |
| CD-RAC | 0.0663 | 0.127 | 0.28 |
| xTB | 0.0410 | 0.078 | 0.70 |
| B3LYP DFT | 0.0600 | 0.115 | 0.32 |
| ωPBEh DFT | 0.0464 | 0.089 | 0.59 |

**Table S16.** The MAE, scaled MAE, and $R^2$ test set performance of the eight grouped split ANNs predicting for lifetime.

|  | MAE (μs) | scaled MAE | $R^2$ |
|---|---|---|---|
| Morgan | 1.1760 | 0.049 | 0.06 |
| Dice | 1.0120 | 0.043 | 0.36 |
| RAC | 0.9960 | 0.042 | 0.22 |
| ligand-only RAC | 1.1759 | 0.049 | 0.26 |
| CD-RAC | 1.4060 | 0.059 | -0.05 |
| xTB | 1.0730 | 0.045 | 0.21 |
| B3LYP DFT | 1.0149 | 0.043 | 0.26 |
| ωPBEh DFT | 1.3272 | 0.056 | 0.06 |

**Table S17.** The MAE, scaled MAE, and $R^2$ test set performance of the eight grouped split ANNs predicting for spectral integral.

|  | MAE (photon counts) | scaled MAE | $R^2$ |
|---|---|---|---|
| Morgan | 3.07E+05 | 0.127 | 0.23 |
| Dice | 5.56E+05 | 0.229 | -1.51 |
| RAC | 4.20E+05 | 0.173 | -0.48 |
| ligand-only RAC | 5.50E+05 | 0.226 | -1.16 |
| CD-RAC | 4.34E+05 | 0.178 | -0.66 |
| xTB | 3.36E+05 | 0.138 | 0.08 |
| B3LYP DFT | 3.44E+05 | 0.141 | -0.04 |
| ωPBEh DFT | 3.81E+05 | 0.157 | -0.23 |

**Table S18.** The percent change in test set MAE for each ANN feature set from the random split to the grouped split.

|  | $Em_{50/50}$ | lifetime | spectral integral |
|---|---|---|---|
| xTB | 95 | 26 | 102 |
| Morgan | 220 | 59 | 137 |
| RAC | 99 | -4 | 115 |
| B3LYP DFT | 107 | 0 | 90 |
| ωPBEh DFT | 63 | 39 | 99 |
| ligand-only RAC | 194 | 20 | 200 |
| Dice | 348 | 34 | 355 |
| CD-RAC | 185 | 52 | 156 |
| average | 164 | 28 | 157 |

**Table S19.** The ranking of grouped split ANNs using different feature sets in predicting for each of the target properties. The ranking is on the basis of test set MAE, and a rank of 1 indicates the corresponding feature set led to the ANN with the lowest MAE for the specified target property.

|  | $Em_{50/50}$ rank | lifetime rank | spectral integral rank | mean rank |
|---|---|---|---|---|
| xTB | 1 | 4 | 2 | 2.33 |
| Morgan | 3 | 6 | 1 | 3.33 |
| RAC | 4 | 1 | 5 | 3.33 |
| B3LYP DFT | 5 | 3 | 3 | 3.67 |
| ωPBEh DFT | 2 | 7 | 4 | 4.33 |
| ligand-only RAC | 6 | 5 | 7 | 6.00 |
| Dice | 8 | 2 | 8 | 6.00 |
| CD-RAC | 7 | 8 | 6 | 7.00 |

**Table S20.** The random split test set performance of a default sklearn linear model with L2 regularization, a sklearn random forest regressor, and an ANN in predicting for each target property. All models were provided with xTB features.

|  |  | MAE | $R^2$ |
|---|---|---|---|
| $Em_{50/50}$ | linear | 0.0683 | 0.27 |
|  | random forest | 0.0317 | 0.80 |
|  | ANN | 0.0210 | 0.91 |
| lifetime | linear | 1.2254 | 0.15 |
|  | random forest | 0.8148 | 0.46 |
|  | ANN | 0.8495 | 0.38 |
| spectral integral | linear | 3.68E+05 | 0.26 |
|  | random forest | 1.80E+05 | 0.76 |
|  | ANN | 1.67E+05 | 0.79 |

**Table S21.** The HLS and CSD ligands that appear most often in hypothetical iridium complexes with random split ANN-predicted properties at the high and low extremes, out of the 3,598 hypothetical complexes considered. The six letter identifiers are CSD refcodes. Only complexes within the UQ cutoffs are considered.

| | | ligand | appearances |
|---|---|---|---|
| $Em_{50/50}$ | top 10$^{th}$ percentile | FEQSEB **NN** | 31 |
| | | YUWWOD **NN** | 29 |
| | | RASGAV **NN** | 25 |
| | bottom 10$^{th}$ percentile | TUZHEE **NN** | 27 |
| | | LISMIK **NN** | 25 |
| | | TOTPAW **NN** | 23 |
| lifetime | top 10$^{th}$ percentile | RASGAV **NN** | 49 |
| | | FEQSEB **NN** | 43 |
| | | MIMYEO **NN** | 41 |
| | bottom 10$^{th}$ percentile | TOTPAW **NN** | 40 |
| | | OVALEE **NN** | 37 |
| | | MAXWIS **CN** | 22 |
| spectral integral | top 10$^{th}$ percentile | YUWWOD **NN** | 39 |
| | | ACUROI **NN** | 35 |
| | | **NN33** | 29 |
| | bottom 10$^{th}$ percentile | KAFPEO **NN** | 54 |
| | | **NN37** | 27 |
| | | QEQVOA **NN** | 26 |

**Table S22.** The ligands present in the hypothetical complexes with the most extreme random split ANN-predicted properties. The six letter identifiers are CSD refcodes. 232 out of 3,598 hypothetical complexes have a predicted spectral integral of zero, preventing the identification of just three extreme complexes on the low end of spectral integral. Only complexes within the UQ cutoffs are considered, and for the case of $Em_{50/50}$ and lifetime only complexes that are predicted to be bright (i.e. spectral integral greater than $1 \times 10^5$ counts) are considered.

|  |  | **CN** ligand | **NN** ligand | predicted value |
|---|---|---|---|---|
| $Em_{50/50}$ | highest | RADTEZ **CN** | **NN43** | 2.4669 |
|  |  | RADTEZ **CN** | **NN3** | 2.4657 |
|  |  | RADTEZ **CN** | YUWWOD **NN** | 2.4647 |
|  | lowest | **CN2** | LEZJAD **NN** | 1.9682 |
|  |  | **CN101** | TUZHEE **NN** | 1.9641 |
|  |  | **CN54** | LEZJAD **NN** | 1.9631 |
| lifetime | highest | **CN95** | TOTPAW **NN** | 10.4054 |
|  |  | **CN95** | YUWWOD **NN** | 10.0016 |
|  |  | **CN67** | RASGAV **NN** | 9.8935 |
|  | lowest | MAXWIS **CN** | **NN1** | 0.0632 |
|  |  | **CN9** | TOTPAW **NN** | 0.0517 |
|  |  | **CN101** | CIDDAX **NN** | 0.0491 |
| spectral integral | highest | CIGKIP **CN** | **NN24** | 2.26E6 |
|  |  | RADTEZ **CN** | **NN20** | 2.20E6 |
|  |  | **CN103** | RASGAV **NN** | 2.19E6 |

**Table S23.** Pearson and Spearman's rank correlation coefficients between experimental properties and B3LYP TDDFT-predicted properties over 26 representative test set complexes.

|  | geometry | Pearson | Spearman's |
|---|---|---|---|
| $Em_{50/50}$ | singlet | 0.81 | 0.86 |
|  | triplet | 0.68 | 0.63 |
| lifetime | singlet | 0.88 | 0.89 |
|  | triplet | 0.67 | 0.44 |

**Table S24.** The 26 experimental complexes evaluated with TDDFT in order to benchmark random split ANN predictions. Here, complexes are represented by the combination of a **CN** and **NN** ligand.

| CN ligand | NN ligand |
|---|---|
| **CN101** | **NN40** |
| **CN101** | **NN41** |
| **CN103** | **NN26** |
| **CN107** | **NN41** |
| **CN109** | **NN40** |
| **CN14** | **NN20** |
| **CN2** | **NN3** |
| **CN28** | **NN41** |
| **CN3** | **NN34** |
| **CN3** | **NN40** |
| **CN31** | **NN33** |
| **CN31** | **NN34** |
| **CN31** | **NN6** |
| **CN38** | **NN27** |
| **CN39** | **NN3** |
| **CN39** | **NN41** |
| **CN54** | **NN1** |
| **CN54** | **NN40** |
| **CN63** | **NN42** |
| **CN69** | **NN27** |
| **CN69** | **NN33** |
| **CN75** | **NN16** |
| **CN77** | **NN3** |
| **CN81** | **NN34** |
| **CN95** | **NN42** |
| **CN95** | **NN8** |

**Table S25.** Pearson and Spearman's rank correlation coefficients between experimental properties and random split ANN-predicted properties over 26 representative test set complexes.

|  | Pearson | Spearman's |
|---|---|---|
| $Em_{50/50}$ | 0.98 | 0.98 |
| lifetime | 0.54 | 0.82 |

**Table S26.** Random split ANN predictions on the bright complexes in the test data with the longest lifetimes.

| CN ligand | NN ligand | experimental lifetime (μs) | ANN-predicted lifetime (μs) |
|---|---|---|---|
| **CN101** | **NN40** | 23.84 | 3.58 |
| **CN101** | **NN41** | 12.74 | 3.73 |
| **CN95** | **NN37** | 12.38 | 7.59 |
| **CN95** | **NN8** | 11.39 | 9.49 |
| **CN95** | **NN3** | 11.23 | 10.26 |
| **CN28** | **NN40** | 10.9 | 2.17 |
| **CN28** | **NN41** | 9.99 | 1.83 |
| **CN95** | **NN42** | 5.95 | 3.13 |
| **CN11** | **NN20** | 5.2 | 3.93 |
| **CN34** | **NN34** | 5.11 | 6.30 |
| **CN38** | **NN27** | 4.66 | 2.52 |
| **CN29** | **NN33** | 4.58 | 2.87 |
| **CN42** | **NN40** | 4.5 | 2.70 |
| **CN35** | **NN20** | 4.38 | 2.72 |
| **CN38** | **NN26** | 4.33 | 3.49 |

**Table S27.** The 21 hypothetical complexes evaluated with TDDFT in order to benchmark random split ANN predictions. Here, complexes are represented by the combination of a **CN** and **NN** ligand.

| CN ligand | NN ligand |
|---|---|
| **CN101** | CIDDAX_eq_lig_0 |
| **CN101** | TUZHEE_eq_lig_2 |
| **CN107** | MUTMOF_eq_lig_2 |
| **CN2** | LEZJAD_eq_lig_2 |
| **CN54** | LEZJAD_eq_lig_2 |
| **CN67** | RASGAV_eq_lig_2 |
| **CN67** | TOTPAW_eq_lig_0 |
| **CN76** | REWDII_eq_lig_0 |
| **CN79** | GEMXAZ_eq_lig_2 |
| **CN9** | TOTPAW_eq_lig_0 |
| **CN95** | MIMYEO_eq_lig_0 |
| **CN95** | TOTPAW_eq_lig_0 |
| **CN95** | YUWWOD_eq_lig_2 |
| HALLEO_ax_lig_0 | FEQSEB_eq_lig_2 |
| MAXWIS_ax_lig_0 | **NN1** |
| OJUSET_ax_lig_0 | TUZHEE_eq_lig_2 |
| RADTEZ_ax_lig_0 | **NN3** |
| RADTEZ_ax_lig_0 | **NN33** |
| RADTEZ_ax_lig_0 | YUWWOD_eq_lig_2 |
| RANGOE_ax_lig_0 | **NN47** |
| SUHLOZ_ax_lig_0 | **NN34** |

**Table S28.** Pearson and Spearman's rank correlation coefficients between random split ANN-predicted properties and B3LYP TDDFT-predicted properties. Type indicates whether the correlation coefficients are evaluated over the 26 representative phosphors from the experimental dataset or over the 21 representative hypothetical phosphors.

|  | geometry | type | Pearson | Spearman's |
|---|---|---|---|---|
| $Em_{50/50}$ | singlet | experimental | 0.81 | 0.89 |
|  | singlet | hypothetical | 0.67 | 0.79 |
|  | triplet | experimental | 0.67 | 0.65 |
|  | triplet | hypothetical | 0.65 | 0.75 |
| lifetime | singlet | experimental | 0.71 | 0.85 |
|  | singlet | hypothetical | 0.05 | 0.32 |
|  | triplet | experimental | 0.79 | 0.54 |
|  | triplet | hypothetical | -0.21 | -0.13 |

**Table S29.** Hyperopt-selected hyperparameters for the best-performing random split ANNs. The lists in the architecture row indicate the number of hidden layers and the number of nodes in each layer. Learning rate, beta1, and decay all affect the Adam optimizer that we use when training our ANNs. bypass indicates the presence of layers that concatenate inputs from the input layer and non-adjacent ANN layers. res indicates the presence of layers that add inputs from non-adjacent ANN layers.

|  | Dice feature set | | | xTB feature set | | |
|---|---|---|---|---|---|---|
|  | $Em_{50/50}$ | lifetime | spectral integral | $Em_{50/50}$ | lifetime | spectral integral |
| architecture | (256, 256, 256) | (512, 512, 512) | (512, 512, 512) | (512, 512, 512) | (256, 256, 256) | (256, 256, 256) |
| learning rate | 0.00069 | 0.00089 | 0.00020 | 0.00085 | 0.00080 | 0.00068 |
| beta1 | 0.880 | 0.811 | 0.982 | 0.827 | 0.848 | 0.873 |
| decay | 0.00179 | 0.00321 | 0.00039 | 0.00124 | 0.00035 | 0.00246 |
| L2 regularization | 0.0105 | 0.0051 | 0.0333 | 0.0044 | 0.0041 | 0.0654 |
| dropout rate | 0.354 | 0.356 | 0.256 | 0.203 | 0.294 | 0.054 |
| batch size | 32 | 16 | 16 | 256 | 128 | 256 |
| epochs | 1122 | 1052 | 1085 | 2000 | 1152 | 2000 |
| bypass | True | False | True | True | True | True |
| res | True | True | True | True | False | False |

**Table S30.** The most different HLS ligands as measured by Dice similarities of Morgan fingerprints. For each **CN** (**NN**) ligand in the HLS, its similarity with each other **CN** (**NN**) ligand in the HLS was taken. For each ligand, these similarities were then averaged to yield a metric for the similarity of that ligand to the rest of the HLS. The lowest averages of similarities are reported below. The three most different **CN** ligands and the two most different **NN** ligands were used to form an out-of-distribution grouped split test set.

| CN ligand | average Dice similarity | NN ligand | average Dice similarity |
|---|---|---|---|
| *CN103* | 0.274 | *NN43* | 0.210 |
| *CN104* | 0.290 | *NN20* | 0.251 |
| *CN21* | 0.300 | **NN21** | 0.274 |
| **CN108** | 0.305 | **NN24** | 0.277 |
| **CN109** | 0.312 | **NN34** | 0.282 |

**Table S31.** Statistics on structure attrition for hit CSD complexes used to identify **CN** and **NN** ligands outside of the HLS. There were six reasons why any CSD complex could be eliminated from consideration: the presence of multiple iridium atoms; the absence of any iridium atoms due to the presence of another molecule larger than the iridium complex in the CSD entry (combined with the "Export largest molecule only" option, as described in Text S1); the incomplete removal of solvent or counterions leading to index errors; the presence of non-bidentate ligands such that a complex was not 2-2-2, i.e. did not have three bidentate ligands in octahedral geometry; refcode duplicates such as HULVEQ and HULVEQ01; and the presence of a **CC** ligand.

| Starting complexes | Multiple Ir atoms | No Ir atoms | Solvent/ counterion | Not 2-2-2 | Refcode duplicates | Presence of a **CC** ligand | Final complexes |
|---|---|---|---|---|---|---|---|
| 700 | 33 | 4 | 7 | 32 | 5 | 1 | 618 |